\documentclass[12pt,draftclsnofoot,onecolumn]{IEEEtran}
\usepackage{cite}
\usepackage{amsmath,amssymb,amsfonts,bm}
\usepackage{graphicx}
\usepackage{textcomp}
\usepackage[ruled,linesnumbered]{algorithm2e}
\usepackage{epsfig,amsbsy,bbm}
\usepackage[tight]{subfigure}
\usepackage{multirow}

\usepackage{epstopdf}
\usepackage{color}
\usepackage{authblk}

\begin{document}
\title{
Semi-Data-Aided Channel Estimation for \\ MIMO Systems via Reinforcement Learning
}

\author{Tae-Kyoung Kim, Yo-Seb Jeon, Jun Li, Nima Tavangaran, and H. Vincent Poor
\thanks{This paper was presented in part at the 2020 IEEE International Conference on Communications (ICC) \cite{Jeon2020-1}.}
\thanks{Tae-Kyoung Kim is with the Department of Electronic Engineering, Gachon University, Seongnam 13120, Republic of Korea (e-mail: tk415kim@gmail.com).}
\thanks{Yo-Seb Jeon is with the Department of Electrical Engineering, POSTECH, Pohang 37673, Republic of Korea (e-mail: yoseb.jeon@postech.ac.kr).}
\thanks{Jun Li is with the School of Electronic and Optical Engineering, Nanjing University of Science and Technology, Nanjing 210094, China. He is also with the Department of Software Engineering, Institute of Cybernetics, National Research Tomsk Polytechnic University, Tomsk 634050, Russia (e-mail: jun.li@njust.edu.cn).}
\thanks{Nima Tavangaran and H. Vincent Poor are with the Department of Electrical and Computer Engineering, Princeton University, Princeton, NJ 08544 USA (e-mail: nimat@princeton.edu; poor@princeton.edu).}
}

\maketitle 
\vspace{-15mm}

\setlength\arraycolsep{2pt}
\newcommand{\argmax}{\operatornamewithlimits{argmax}}
\newcommand{\argmin}{\operatornamewithlimits{argmin}}

\newtheorem{defn}{Definition}
\newtheorem{thm}{Theorem}
\newtheorem{lem}{Lemma}
\newtheorem{cor}{Corollary}
\newtheorem{prop}{Proposition}
\newtheorem{rem}{Remark}
\newtheorem{pro}{Property}
\newtheorem{app}{Appendix}

\begin{abstract}
Data-aided channel estimation is a promising solution to improve channel estimation accuracy by exploiting data symbols as pilot signals for updating an initial channel estimate. In this paper, we propose a semi-data-aided channel estimator for multiple-input multiple-output communication systems. Our strategy is to leverage reinforcement learning (RL) for selecting reliable detected symbols  among the symbols in the first part of transmitted data block. This strategy facilitates an update of the channel estimate before the end of data block transmission and therefore achieves a significant reduction in communication latency compared to conventional data-aided channel estimation approaches. Towards this end, we first define a Markov decision process (MDP) which sequentially decides whether to use each detected symbol as an additional pilot signal. We then develop an RL algorithm to efficiently find the best policy of the MDP based on a Monte Carlo tree search approach. In this algorithm, we exploit the a-posteriori probability for approximating both the optimal future actions and the corresponding state transitions of the MDP and derive a closed-form expression for the best policy. Simulation results demonstrate that the proposed channel estimator effectively mitigates both channel estimation error and detection performance loss caused by insufficient pilot signals.
\end{abstract}
\vspace{-5mm}

\section{Introduction}

Multiple-input multiple-output (MIMO) communication is one of the core technologies in modern wireless standards.
The use of multiple antennas significantly improves both the capacity and the reliability of wireless systems by providing spatial multiplexing and diversity gains \cite{Foschini1996,Telatar1999,Zheng2003}.
A key requirement to enjoy these benefits is accurate channel state information (CSI) at both transmitter and receiver.
For example, the capacity of MIMO communication systems increases linearly with the number of either transmit or receive antennas under the premise that perfect CSI is available at both the transmitter and receiver \cite{Foschini1996,Telatar1999}.

To obtain accurate CSI at the receiver (CSIR), various channel estimation techniques have been developed for MIMO communication systems  \cite{Simeone2004,Kim2014,Biguesh2006,Ozdemir2007,Dowler2003,Zhao2008,Huang2018,Liu2012,Kim2012,Verenzuela2020,Ma2014,Park2015,Park2017}.
One of the most popular and widely adopted technique is pilot-aided channel estimation \cite{Simeone2004,Kim2014,Biguesh2006,Ozdemir2007}.
The fundamental idea of this technique is to send pilot signals, known as a priori at the receiver, and then to estimate the CSI from received signals observed during pilot transmission. 
A representative example of this technique is the least-squares (LS) channel estimator that minimizes the sum of squared errors in the estimated CSIR \cite{Biguesh2006,Ozdemir2007}. 
Another example is the linear minimum-mean-squared-error (LMMSE) channel estimator which is a linear estimator that minimizes the mean-squared-error (MSE) of the estimated CSIR based on the first-order and the second-order channel statistics \cite{Biguesh2006,Ozdemir2007}. 
The accuracy of the CSIR obtained from pilot-aided channel estimation improves with the number of the pilot signals available in a communication system.
In addition, the larger the number of spatially multiplexed data streams utilized in MIMO systems, the larger the number of pilot signals required for accurate CSIR.
Despite this requirement,  in practical MIMO communication systems, only a small portion of radio resources are allocated for pilot transmission, while most of the radio resources are allocated for transmitting data (non-pilot) signals. 

Data-aided channel estimation is a promising solution to overcome the limitation of pilot-aided channel estimation due to an insufficient number of pilot signals \cite{Dowler2003,Liu2012,Kim2012,Verenzuela2020,Ma2014,Huang2018,Zhao2008,Park2015,Park2017}.
The basic strategy of the data-aided channel estimation is to exploit data symbols as additional pilot signals to update an initial channel estimate obtained from pilot-aided channel estimation.
This strategy allows the receiver to enjoy the effect of increasing the number of pilot signals and therefore has a potential to provide more accurate CSIR compared to the pilot-aided channel estimation without sacrificing radio resource for data transmission. 
A non-iterative data-aided channel estimation was first investigated in \cite{Dowler2003}.
In this method, data symbols are reconstructed by properly encoding and modulating the outputs of channel decoder, so that the reconstructed data symbols are utilized as pilot signals for channel estimation. 
The performance of this method, however, is degraded under the presence of decoding error which leads to the mismatch between the reconstructed and transmitted data symbols.
To resolve this problem, iterative data-aided channel estimation has been studied in \cite{Zhao2008,Ma2014,Park2015,Huang2018,Park2017}, which iteratively performs channel estimation and data detection to mitigate both channel estimation and decoding errors.
In \cite{Zhao2008}, an iterative turbo channel estimation technique was developed in which soft-decision symbols are utilized as pilot signals at each iteration.
A similar iterative approach was also developed in \cite{Park2015} by selectively utilizing soft-decision symbols as pilot signals according to an MSE-based criterion.
The common limitation of these iterative data-aided channel estimators is that they increase not only the computational complexity of receive processing, but also communication latency.

Recently, deep-learning-based channel estimation has also drawn increasing attention in order to circumvent the limitation of pilot-aided channel estimation \cite{Neumann2018,Zhang2021,Chun2019,Yang2019,He2018,Qiang2020,Wei2019,He2020,Ma2021}. 
A basic idea of this technique is to learn a channel from training samples, each of which describes the input-output relation of a communication system.
The most prominent feature of the deep-learning-based channel estimation is that it can be readily incorporated into complicated communication systems, e.g., massive MIMO, millimeter-wave, and doubly-selective channels \cite{Chun2019,Yang2019,He2018}.
The use of deep learning, however, requires a huge training set to optimize  neural networks and therefore increases both computational complexity and communication latency.
To resolve this drawback, a model-driven deep learning approach was studied in \cite{Qiang2020,Wei2019,He2020,Ma2021}.
This approach effectively reduces the size of training set by learning only the parameters of a model for estimating the channel. 
Specifically, a joint optimization with data detection and channel estimation was introduced in \cite{He2020} based on a Bayesian model. 
A similar channel estimation method for millimeter-wave MIMO systems was introduced in \cite{Ma2021}.
Although these model-driven channel estimators effectively mitigate the limitation of the deep learning-based channel estimation, the use of deep learning still brings non-negligible computational complexity and communication latency that may not be affordable in practical systems. 

This paper presents a new type of data-aided channel estimation for MIMO communication systems, referred to as {\em semi-data-aided} channel estimation, which reduces communication latency caused by iterative data-aided channel estimation.
The basic strategy of the presented channel estimator is to leverage reinforcement learning (RL) for selecting reliable detected symbol vectors only among the symbols in the first part of data block. 
The most prominent feature of the presented channel estimator is that it does not utilize the channel decoder outputs and therefore facilitates an early update of a channel estimate even before the end of data block transmission.
Simulation results demonstrate that the presented channel estimator effectively mitigate both channel estimation error and detection performance loss caused by insufficient pilot signals.
%
The major contributions of this paper are summarized as follows:
\begin{itemize}
    \item We present a Markov decision process (MDP) to sequentially determine the best selection of detected symbol vectors for minimizing the MSE of the semi-data-aided channel estimation.
    To this end, we adopt a binary action that indicates whether to exploit each detected symbol vector as an additional pilot signal, while defining a reward function as the MSE reduction of the channel estimate.
    With this MDP, we successfully formulate a symbol vector selection problem for the semi-data-aided channel estimation as a sequential decision-making problem that can be efficiently solved via RL. 
    

     \item We propose a novel RL algorithm to efficiently find the best policy of the presented MDP. 
     The underlying challenge is that the state transition of the presented MDP is unknown at the receiver due to the lack of knowledge of transmitted symbol vectors. 
     In the proposed algorithm, we tackle this challenge by leveraging a Monte Carlo tree search (MCTS) approach in \cite{Sutton,Browne2012,Vodopivec2017} which looks ahead the rewards of near-future actions, while approximating the rewards of distant-future actions via Monte Carlo simulations. 
     We modify the original MCTS approach by exploiting a-posteriori probability (APP), computed from data detection, for approximating both the optimal future actions and the corresponding state transitions of the MDP.   
     The most prominent advantage of the proposed RL algorithm is that the best policy for each state has a closed-form expression that can be readily computed at the receiver.

    \item We present two additional strategies for enhancing the advantages of  the semi-data-aided channel estimation operating with the proposed RL algorithm. 
    In the first strategy, we develop a low-complexity policy that approximates the optimal policy of the presented MDP based on Monte Carlo sampling. 
    Utilizing this new policy, we further reduce the computational complexity required in the proposed RL algorithm.
    In the second strategy, we utilize an updated channel estimate for re-detecting the symbol vectors that are not selected by the proposed RL algorithm. Utilizing this strategy, we further improve data detection performance when employing the semi-data-aided channel estimation, without a significant increase in the computational complexity. 

    
    \item In simulations, we evaluate the normalized MSE (NMSE) and block-error-rate (BLER) of the proposed channel estimator for a coded MIMO communication system. 
    Our simulation results demonstrate that the proposed channel estimator significantly reduces the NMSE in channel estimation, while improving the BLER of the system, compared to conventional pilot-aided channel estimation.
    It is also shown that the proposed RL algorithm effectively selects detected symbol vectors that can improve the performance of the semi-data-aided channel estimation. 
    We also investigate the robustness of the proposed channel estimator in time-varying channels and demonstrate that the proposed channel estimator reduces performance degradation in time-varying environment by tracking temporal channel variations during data transmission. 
  
\end{itemize}

An RL algorithm for optimizing the symbol vector selection of data-aided channel estimation was first introduced in our prior work \cite{Jeon2020-1}.
In this algorithm, the optimal policy of the MDP is derived under a simplistic assumption that underestimates the effect of future actions and rewards. 
In this paper, we generalize the RL algorithm in \cite{Jeon2020-1} by employing the MCTS approach which provides a more accurate evaluation of the effect of the future actions and rewards.
In addition to this major change, we newly introduce the semi-data-aided channel estimation strategy to further reduce the delay required for updating the channel estimate and also introduce the data re-detection strategy to improve detection performance after the symbol vector selection. 


The remainder of this paper is organized as follows.
Section II introduces system model and preliminaries considered in this paper.
In Section III, we formulate an optimization problem that adaptively selects the detected symbols for the semi-data-aided channel estimator.
An efficient RL algorithm to solve the optimization problem is proposed in Section IV.
Simulation results are presented in Section V to verify the effectiveness of the proposed channel estimator.
The conclusion is finally presented in Section VI.

\subsubsection*{Notation}
Matrices $\mathbf{0}_{m}$ and $\mathbf{I}_{m}$ represent the $m{\times} m$ all-zero matrix and the $m\times m$ identity matrix, respectively.
Superscripts $(\cdot)^T$ and $(\cdot)^H$ denote the transpose and the conjugate transpose, respectively.
Operators $\mathbb{E}(\cdot)$, $\mathbb{P}(\cdot)$, $| \cdot |$, and $\| \cdot \|_{\rm F}$ denote the expectation of a random variable, the probability of an event, the cardinality of a set, and the Frobenius norm, respectively.
$(\cdot)^{-1}$ denotes the inverse operation.
The set $\mathbb{C}$ represents the set of complex numbers.

\section{System Model and Preliminaries}
In this section, we introduce a MIMO communication system considered in this work. 
The LMMSE channel estimator and the maximum-a-posteriori-probability (MAP) data detector are presented for the considered system.
We then describe the challenge of the LMMSE channel estimator to achieve the optimal performance.

\begin{figure}
	\centering
	\epsfig{file=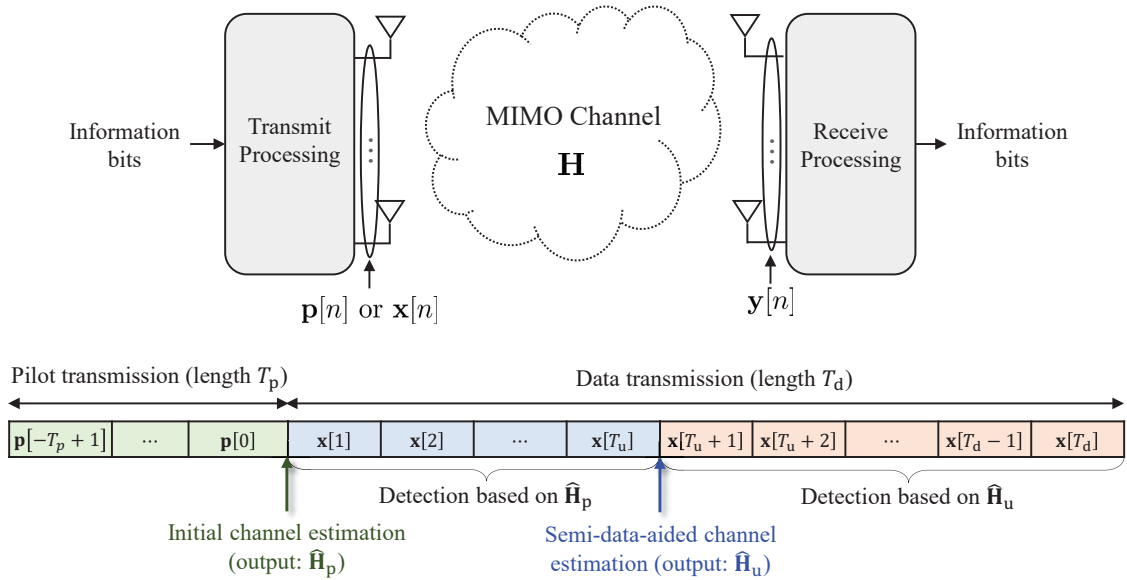, width=15cm}
	\caption{A MIMO communication system in which a transmitter equipped with $N_{\rm{tx}}$ antennas communicates with a receiver equipped with $N_{\rm{rx}}$ antennas. A transmission frame consists of a pilot block with length $T_{\rm{p}}$ followed by a data block with length $T_{\rm{d}}$. The data block consists of two parts: The lengths of the first part and the second part are $T_{\rm u}$ and $T_{\rm d}-T_{\rm u}$, respectively.}
	\label{Fig:System}
\end{figure}

\subsection{System model}
We consider a MIMO communication system in which a transmitter equipped with $N_{\rm{tx}}$ antennas communicates with a receiver equipped with $N_{\rm{rx}}$ antennas, as illustrated in Fig.~\ref{Fig:System}.
We model the wireless channel of the considered system as a frequency-flat Rayleigh fading channel denoted by $\mathbf{H}\in \mathbb{C}^{N_{\rm{rx}}\times N_{\rm{tx}}}$ where the entries of ${\bf H}$ are independent and identically distributed (i.i.d.) random variables with the distribution of $\mathcal{CN}(0,1)$.  
We assume a block fading channel in which the entries of $\mathbf{H}$ keep constant during a transmission frame.

A transmission frame consists of a pilot block with length $T_{\rm{p}}$ followed by a data block with length $T_{\rm{d}}$, as illustrated in Fig.~\ref{Fig:System}.
A set of time slot indices associated with the pilot block and the data block is denoted by $\mathcal{T}_{\rm p}=\{-T_{\rm p}+1,\ldots,0\}$ and $\mathcal{T}_{\rm d}=\{1,\ldots,T_{\rm{d}}\}$, respectively. 
Let $\mathbf{p}[n]\in\mathbb{C}^{N_{\rm{tx}}}$ be the pilot signal sent at time slot $n$ such that $\mathbb{E}\left[\|\mathbf{p}[n]\|^{2}\right]=N_{\rm{tx}}$.
Then the received signal at time slot $n\in\mathcal{T}_{\rm p}$ is given by
\begin{align}\label{II.A.1}
	\mathbf{y}[n]= \mathbf{H}\mathbf{p}[n]+ {\bf z}[n],
\end{align}
where $\mathbf{z}[n]\sim\mathcal{CN}\left(\mathbf{0}_{N_{\rm{rx}}},\sigma^2\mathbf{I}_{N_{\rm{rx}}}\right)$ is a circularly symmetric complex Gaussian noise vector at time slot $n$.
For the data transmission, the transmitter generates data symbol vectors after symbol mapping of information bits.
Let $\mathbf{x}[n]\in \mathcal{X}^{N_{\rm{tx}}}$ be the data symbol vector sent at time slot $n\in \mathcal{T}_{\rm d}$, where $\mathcal{X}$ is a constellation set such that $\mathbb{E}\left[\|\mathbf{x}[n]\|^{2}\right]=N_{\rm{tx}}$.
Then the received signal at time slot $n\in\mathcal{T}_{\rm d}$ during the data transmission is given by
\begin{align}\label{II.A.2}
\mathbf{y}[n] &= \left[y_{1}[n], \cdots ,y_{N_{\rm{rx}}}[n] \right]^{T} = \mathbf{H}\mathbf{x}[n]+\mathbf{z}[n].
\end{align}

\subsection{LMMSE Channel Estimator}

The LMMSE channel estimator is a linear estimator that minimizes the MSE of a channel estimate.
This method has been widely adopted in wireless communication systems as it provides a good trade-off between estimation accuracy and computational complexity \cite{Biguesh2006,Ozdemir2007}.
Let ${\bf Y}_{\rm{p}}$ a matrix that concatenates received signals observed during the pilot transmission. 
From \eqref{II.A.1}, ${\bf Y}_{\rm{p}}$ is expressed as
\begin{align}\label{eq:Received_pilot}
	{\bf Y}_{\rm{p}} &= \big[{\bf y}[-T_{\rm p}+1],\cdots,{\bf y}[0] \big]
	= \mathbf{H}{\bf P}+ {\bf Z}_{\rm{p}},
\end{align}
where $\mathbf{P}  =\big[\mathbf{p}[-T_{\rm p}+1],\cdots,\mathbf{p}[0]\big]$, and $\mathbf{Z}_{\rm{p}}  =\big[\mathbf{z}[-T_{\rm p}+1],\cdots,\mathbf{z}[0]\big]$.
From \eqref{eq:Received_pilot}, the LMMSE channel estimator is given by
\begin{align}\label{II.B.1}
\mathbf{W}_{\rm{LMMSE}} &= \argmin_{{\bf W}\in \mathbb{C}^{ T_{\rm{p}} \times N_{\rm{tx}}}} \mathbb{E} \left[\| {\bf Y}_{\rm p}{\bf W} - {\bf H}\|_{\rm F}^{2}\right]\nonumber\\
&= \mathbf{P}^{H} \big(\mathbf{P}\mathbf{P}^{H}+\sigma^{2}\mathbf{I}_{N_{\rm{tx}}}\big)^{-1},
\end{align}
where the expectation is taken with respect to channel and noise distributions.
Consequently, the LMMSE channel estimate is computed as
\begin{align}\label{II.B.2}
    \hat{\bf H}_{\rm p} &=  {\bf Y}_{\rm p} \mathbf{P}^{H}
    \big(\mathbf{P}\mathbf{P}^{H}+\sigma^{2}\mathbf{I}_{N_{\rm{tx}}}\big)^{-1}.
\end{align}
If the entries of ${\bf H}$ are i.i.d. with $\mathcal{CN}(0,1)$, the MSE of the LMMSE channel estimate is computed as 
\begin{align}\label{eq:MSE_initial}
    \mathbb{E} \big[\| \hat{\bf H}_{\rm p} - {\bf H}\|_{\rm F}^{2}\big]
    &=  N_{\rm rx} {\rm Tr} \big[ \mathbb{E} \big[ (\hat{\bf h}_{{\rm p},r}^{H} - {\bf h}_r^{H}) (\hat{\bf h}_{{\rm p},r} - {\bf h}_r) \big] \big] \nonumber \\
    &= N_{\rm rx} \sigma^2 {\rm Tr} \big[ \big( {\bf P}{\bf P}^{H} + \sigma^2 {\bf I}_{N_{\rm tx}}\big)^{-1} \big],
\end{align}
where $\hat{\bf h}_{{\rm p},r}$ and ${\bf h}_r$ are the $r$-th row of $\hat{\bf H}_{\rm p}$ and ${\bf H}$, respectively. 
As can be seen from \eqref{eq:MSE_initial}, the MSE of the LMMSE channel estimate decreases with the number of the pilot signals $T_{\rm p}$.

\subsection{Maximum-A-Posteriori-Probability (MAP) Data Detector}
In this work, we assume that the receiver employs the MAP data detection method which finds the symbol vector with the maximum APP for a received signal. 
This method is optimal in terms of minimizing detection error probability and therefore has a potential to maximize the performance of a channel estimator presented in Sec. IV. 
Nevertheless, as will be discussed later, the applicability of the presented channel estimator is not limited to the MAP data detection method.

Let $\mathbf{x}_{k}$ be a vector in $\mathcal{X}^{N_{\rm{tx}}}$ with $k\in\mathcal{K}=\{1,\ldots,K\}$ where $K=|\mathcal{X}|^{N_{\rm{tx}}}$.
The APP of the event $\{\mathbf{x}[n]=\mathbf{x}_{k}\}$ for the given received signal $\mathbf{y}[n]$ is expressed as
\begin{align}\label{II.C.1}
{\theta}_{k}[n] 
&\triangleq \mathbb{P}\left[\mathbf{x}[n]=\mathbf{x}_{k}|\mathbf{y}[n]\right] \nonumber \\
&= \frac{ \mathbb{P}\left[\mathbf{y}[n]|\mathbf{x}[n]=\mathbf{x}_{k}\right] \mathbb{P}\left[\mathbf{x}[n]=\mathbf{x}_{k}\right]}{\sum_{j\in\mathcal{K}}\mathbb{P}\left[\mathbf{y}[n]|\mathbf{x}[n]=\mathbf{x}_{j}\right] \mathbb{P}\left[\mathbf{x}[n]=\mathbf{x}_{j}\right]} \nonumber \\
&\overset{(a)}{=} \frac{ \mathbb{P}\left[\mathbf{y}[n]|\mathbf{x}[n]=\mathbf{x}_{k}\right]}{\sum_{j\in\mathcal{K}}\mathbb{P}\left[\mathbf{y}[n]|\mathbf{x}[n]=\mathbf{x}_{j}\right] },
\end{align}
where the equality (a) holds when the probability of transmitting each symbol vector is equal (i.e., $\mathbb{P}\left[\mathbf{x}[n]=\mathbf{x}_{k}\right]=\frac{1}{K}$, $\forall k \in \mathcal{K}$).
Since $\mathbf{z}[n]\sim\mathcal{CN}\left(\mathbf{0}_{N_{\rm{rx}}},\sigma^2\mathbf{I}_{N_{\rm{rx}}}\right)$, the probability $\mathbb{P}\left[\mathbf{y}[n]|\mathbf{x}[n]=\mathbf{x}_{k}\right]$ in \eqref{II.C.1} is given by
\begin{align}\label{II.C.2}
\mathbb{P}\left[\mathbf{y}[n]|\mathbf{x}[n]=\mathbf{x}_{k}\right]
&=\frac{1}{\left(\pi \sigma^{2}\right)^{N_{\rm{rx}}} } \exp\left(-\frac{\|\mathbf{y}[n]-\mathbf{H}\mathbf{x}_{k}\|^{2}}{\sigma^{2}}\right),
\end{align}
for $k\in\mathcal{K}$. This probability is also known as the likelihood function. 
By applying \eqref{II.C.2} into \eqref{II.C.1}, the APP is computed as
\begin{align}\label{eq:APP}
    {\theta}_{k}[n] = \frac{ \exp\big(-\frac{1}{\sigma^{2}}\|\mathbf{y}[n]-{\bf H}\mathbf{x}_{k}\|^{2}\big)}
    {\sum_{j\in\mathcal{K}}\exp\big(-\frac{1}{\sigma^{2}}\|\mathbf{y}[n]-{\bf H}\mathbf{x}_{j}\|^{2}\big)}.
\end{align}
Then the MAP detection rule is given by 
\begin{align}\label{II.C.3}
    \hat{\bf x}[n] = {\bf x}_{\hat{k}_n},~~\text{where}~~\hat{k}_{n} &= \argmax\limits_{k\in\mathcal{K}}~\theta_{k}[n].
\end{align}

In practical communication systems, the receiver cannot compute the exact APP in \eqref{eq:APP} as it requires perfect information of ${\bf H}$. 
As an alternative approach, an approximate APP is utilized for data detection, which is computed based on the MIMO channel estimate $\hat{\bf H}_{\rm p}$ from \eqref{II.B.2} as follows:
\begin{align}\label{eq:APP_approx}
    \hat{\theta}_{k}[n] = \frac{ \exp\big(-\frac{1}{\sigma^{2}}\|\mathbf{y}[n]-\hat{\bf H}_{\rm p}\mathbf{x}_{k}\|^{2}\big)}
    {\sum_{j\in\mathcal{K}}\exp\big(-\frac{1}{\sigma^{2}}\|\mathbf{y}[n]-\hat{\bf H}_{\rm p}\mathbf{x}_{j}\|^{2}\big)}.
\end{align}
Unfortunately, when employing the pilot-aided channel estimation with an insufficient number of the pilot signals, channel estimation error (i.e., $\hat{\bf H}_{\rm p}-{\bf H}$) is inevitable at the receiver, as shown in \eqref{eq:MSE_initial}. 
Because this error leads to a mismatch between the true APP in \eqref{eq:APP} and the approximate APP in \eqref{eq:APP_approx}, the use of the approximate APP results in detection performance degradation. 
Moreover, the degree of the performance degradation increases as the number of the pilot signals, $T_{\rm p}$, reduces. 
To resolve this problem, in the following sections, we will present a novel channel estimation approach that utilizes detected symbol vectors to reduce the channel estimation error caused by  insufficient pilot signals.


\section{Optimization Problem for Semi-Data-Aided Channel Estimation}
Data-aided channel estimation is a well-known approach to reduce channel estimation error when the number of pilot signals is insufficient. 
The fundamental idea of the data-aided channel estimation is to exploit detected symbol vectors as additional pilot signals for updating a channel estimate. 
On the basis of the same idea, in this section, we present a new type of the data-aided channel estimation, referred to as {\em semi-data-aided} channel estimation, which enables fast update of the channel estimate with the selective use of detected symbol vectors.
In what follows, we first elaborate on the basic idea of the semi-data-aided channel estimation and an optimization problem to maximize its performance.
We then reformulate the optimization problem as an MDP in order to adopt RL to solve this problem. 

\subsection{Semi-Data-Aided Channel Estimation}
Our key observation is that not every detected symbol vector is a good candidate for a pilot signal because some detected symbol vectors differ from the transmitted symbol vectors due to data detection error.    
Another important observation is that once the receiver obtains a sufficient number of additional pilot signals, increasing the number of the pilot signals gives no significant improvement in channel estimation accuracy. 
Motivated by these observations, in the semi-data-aided channel estimation, we exploit only the detected symbol vectors that are beneficial for improving the channel estimation accuracy.
Meanwhile, we select these symbol vectors only among the first $T_{\rm u}$ detected symbol vectors, while utilizing the updated channel estimate for detecting the remaining $T_{\rm d}-T_{\rm u}$ symbol vectors, as illustrated in Fig.~\ref{Fig:System}.
We refer to this strategy as a {\em semi-data-aided} channel estimation because it utilizes only a portion of detected symbol vectors, unlike the conventional data-aided channel estimation.
The most prominent advantage of our strategy is that a channel estimate is updated after the transmission of $T_{\rm u}$ symbol vectors; thereby, our strategy significantly reduces the delay required for updating the channel estimate compared to conventional data-aided channel estimation methods that updates the channel estimates after the end of data block transmission (i.e., $T_{\rm d}$ time slots).
Moreover, the semi-data-aided channel estimation does not utilize the outputs of a channel decoder, implying that the repetitions of channel decoding process is not necessary. 
Because of this feature, the computational complexity of the semi-data-aided channel estimation is lower than those of conventional data-aided channel estimation methods which require to repeat the channel decoding process (e.g., \cite{Liu2012, Kim2012, Verenzuela2020, Zhao2008, Park2015, Park2017}).

\subsection{Optimization Problem for Symbol Vector Selection}

A key to the success of the semi-data-aided channel estimation is to optimize the selection of detected symbol vectors so that the accuracy of an updated channel estimate can be maximized. 
A direct optimization of the symbol vector selection, however, is very challenging in practical systems due to the lack of knowledge of transmitted symbol vectors and also due to high computational complexity.
To shed some light on this challenge, we formulate an optimization problem for the symbol vector selection to minimize the error of the updated channel estimate. 
Let ${\bf a} \in \{0,1\}^{T_{\rm u}}$ be a vector whose $n$-th entry indicates whether to utilize the detected symbol vector at time slot $n$, $\hat{\bf x}[n]$, in the semi-data-aided channel estimation. 
If the receiver utilizes only the detected symbol vectors indicated by ${\bf a}$ as additional pilot signals, the LMMSE channel estimate is updated as
\begin{align}
    \hat{\bf H}({\bf a}) 
    = {\bf Y}({\bf a}) {\bf W}_{\rm LMMSE}({\bf a}) =  {\bf Y}({\bf a})  \mathbf{\hat{X}}^{H}({\bf a}) \big(\mathbf{\hat{X}}({\bf a})\mathbf{\hat{X}}^{H}({\bf a}) + \sigma^{2}\mathbf{I}_{N_{\rm{tx}}}\big)^{-1},
\end{align}
where ${\bf Y}({\bf a})= \big[\mathbf{Y}_{\rm{p}}, {\bf y}[l_1({\bf a})],\cdots,{\bf y}[l_{\|{\bf a}\|_0}({\bf a})]\big]$, ${\bf X}({\bf a})= \big[\mathbf{P}, \hat{\bf x}[l_1({\bf a})],\cdots,\hat{\bf x}[l_{\|{\bf a}\|_0}({\bf a})]\big]$, and $l_i({\bf a})$ is the index of the $i$-th nonzero entry in a vector ${\bf a}$. 
Note that $\|{\bf a}\|_0$ is the number of nonzero entries in ${\bf a}$. 
Based on the above notations, a symbol vector selection problem for minimizing the MSE of the updated channel estimate is formulated as
\begin{align}\label{eq:Problem}
    {\bf a}^\star 
    = \underset{{\bf a} \in \{0,1\}^{T_{\rm u}}}{\argmin}~ \mathbb{E}\big[\|\mathbf{\hat{H}}({\bf a}) - \mathbf{H} \|_{\rm F}^{2}\big] = \underset{{\bf a} \in \{0,1\}^{T_{\rm u}}}{\argmin}~ \mathbb{E}\big[\|   {\bf Y}({\bf a}) {\bf W}_{\rm LMMSE}({\bf a})   - \mathbf{H} \|_{\rm F}^{2}\big], 
\end{align}
where the expectation is taken with respect to channel and noise distributions. 
The first key observation is that the distribution of ${\bf Y}({\bf a})$ depends on the transmitted symbol vectors associated with ${\bf a}$; thereby, solving the optimization problem in \eqref{eq:Problem} requires perfect knowledge of the first $T_{\rm u}$ transmitted symbol vectors at the receiver.  
Another important observation is that the number of possible choices for symbol vector selection is given by $2^{T_{\rm u}}$ which exponentially increases with the number of symbol vector candidates. 
These observations reveal that directly solving the problem in \eqref{eq:Problem} is very challenging at the receiver in practical systems.  

\subsection{MDP for Symbol Vector Selection}\label{Sec:MDP}
To circumvent the aforementioned challenge, we reformulate the optimization problem in \eqref{eq:Problem} as an MDP which sequentially decides whether to use each detected symbol vector when its reward is a reduction in channel estimation error.   
In Sec.~IV, we will demonstrate how this MDP allows the receiver to approximately but efficiently solves the original problem in \eqref{eq:Problem} using an RL approach.
Details of our MDP formulation are elaborated below.

\subsubsection{State}
The state set of the MDP associated with time slot $n$ is defined as
\begin{align}\label{III.B.1-1}
 \mathcal{S}_{n} = \big\{&\big(\mathbf{X}_{n},\mathbf{\hat{X}}_{n},{\bf a}_n  \big)|   \nonumber\\
& \mathbf{X}_{n}=\big[\mathbf{P},\mathbf{x}_{j_1},\cdots,\mathbf{x}_{j_{\|{\bf a}_n\|_0}}\big], \mathbf{\hat{X}}_{n}=\big[\mathbf{P},\mathbf{\hat{x}}[l_1({\bf a}_n)],\cdots,\mathbf{\hat{x}}[l_{\|{\bf a}_n\|_0}({\bf a}_n)]\big] , {\bf a}_n \in\{0,1\}^{n-1} \big\},
\end{align}
where $j_i$ is the candidate index for the next transition at the $i$-th nonzero entry in a vector ${\bf a}_{n}$ such that $j_i\in\mathcal{K}$.
In \eqref{III.B.1-1}, ${\bf a}_n$ is the set of the actions until the time slot $n-1$.
If $a_i = 1$, it indicates that the detected symbol vector at time slot $i$ will be exploited as additional pilot signals for the data-aided channel estimation. 
Using this definition, the LMMSE channel estimate obtained at the state $\mathrm{S}_{n} =\big(\mathbf{X}_{n},\mathbf{\hat{X}}_{n},{\bf a}_{n} \big)\in \mathcal{S}_{n}$ is given by
\begin{align}\label{III.B.1-2}
\mathbf{\hat{H}} \left(\mathrm{S}_{n}\right) &=  {\bf Y}(\mathrm{S}_{n})  \mathbf{\hat{X}}_{n}^{H} \big(\mathbf{\hat{X}}_{n}\mathbf{\hat{X}}_{n}^{H}+\sigma^{2}\mathbf{I}_{N_{\rm{tx}}}\big)^{-1},
\end{align}
where ${\bf Y}(\mathrm{S}_{n})= \big[\mathbf{Y}_{\rm{p}}, {\bf y}[l_1({\bf a}_n)],\cdots,{\bf y}[l_{\|{\bf a}_n\|_0}({\bf a}_n)]\big]$.

\subsubsection{Reward Function}

The reward function of the MDP is defined as the MSE reduction of the channel estimate when transiting from the current state to the next state.
Based on this definition, the reward function associated with the state transition from $\mathrm{S}_{n}\in \mathcal{S}_{n}$ to $\mathrm{S}_{n+1}\in \mathcal{S}_{n+1}$ is given by
\begin{align}\label{III.B.4-2}
\mathsf{R}\left(\mathrm{S}_{n},\mathrm{S}_{n+1}\right)&= \frac{1}{N_{\rm rx}} \left\{\mathbb{E}\big[\|\mathbf{\hat{H}}\left(\mathrm{S}_{n}\right) - \mathbf{H} \|_{\rm F}^{2}\big]  - \mathbb{E}\big[\|\mathbf{\hat{H}}\left(\mathrm{S}_{n+1}\right) - \mathbf{H} \|_{\rm F}^{2}\big] \right\}.
\end{align}

\subsubsection{Action}

The action set of the MDP is defined as $\mathcal{A}=\{1,0\}$ which indicates whether to exploit the current detected symbol vector as an additional pilot signal.
For example, the action with $a = 1$ implies that the detected symbol vector will be exploited as the pilot signal.

\subsubsection{State Transition}

From the definitions of the state and action, the current state is updated using the detected symbol vector when $a=1$; otherwise, the current state remains unchanged.
Thus, the state $\mathsf{U}\left(\mathrm{S}_{n}|a\right)\in \mathcal{S}_{n+1}$ that can be transited to the current $\mathrm{S}_{n}=(\mathbf{X}_{n},\mathbf{\hat{X}}_{n},{\bf a}_n)\in\mathcal{S}_{n}$ is given by
\begin{align}\label{III.B.3-2}
\mathsf{U}\left(\mathrm{S}_{n}|a\right)
&=\begin{cases} \big([\mathbf{X}_{n},\mathbf{x}_{k_{n}}],[\mathbf{\hat{X}}_{n},\mathbf{\hat{x}}[n]],[{\bf a}_n,1]\big),& a=1, \\
\big(\mathbf{X}_{n},\mathbf{\hat{X}}_{n},[{\bf a}_n,0]\big),&a=0. \end{cases}
\end{align}

\subsubsection{Optimal Policy}

\begin{figure*}
	\centering
	\epsfig{file=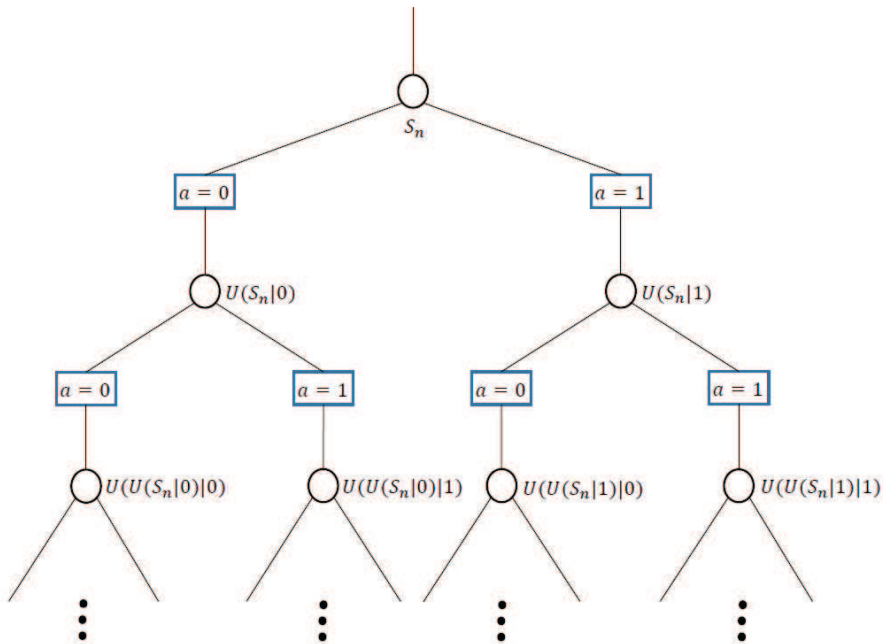, width=12cm}
	\caption{State-action diagram of the original MCTS for $a\in\mathcal{A}$ and $\mathrm{S}_{n}\in\mathcal{S}_{n}$.}
	\label{FigOMDP}
\end{figure*}

The optimal policy of the MDP for a state $\mathrm{S}_{n}\in\mathcal{S}_{n}$ is defined as
\begin{align}\label{III.B.5-1}
\pi^{\star}\left(\mathrm{S}_{n}\right)&=\argmax_{a\in\mathcal{A}} \mathsf{Q}\left(\mathrm{S}_{n},a \right),
\end{align}
where $\mathsf{Q}\left(\mathrm{S}_{n},a \right)$ is the Q-value function that represents the optimal sum of the rewards obtained after taking the action $a\in\mathcal{A}$ at the state $\mathrm{S}_{n}$.
By the definition in \eqref{III.B.3-2}, $\mathsf{Q}\left(\mathrm{S}_{n},a \right)$ can be expressed as
\begin{align}\label{III.B.5-2}
\mathsf{Q}\left(\mathrm{S}_{n},a\right) 
&=\mathsf{R}\left(\mathrm{S}_{n},\mathsf{U}\left(\mathrm{S}_{n}|a\right)\right)+ \gamma\mathsf{V}^{\star}\left(\mathsf{U}\left(\mathrm{S}_{n}|a\right) \right),
\end{align}
where $0\leq \gamma\leq 1$ is a discounting factor, and $\mathsf{V}^{\star}\big(\mathrm{S}_{m}\big)$ is the optimal value function which is the optimal sum of the rewards that can be obtained from the state $\mathrm{S}_{m}\in\mathcal{S}_{m}$ with $m\in\{n+1,\ldots,T_{\rm u}\}$. 
The optimal value function for a state $\mathrm{S}_{m}\in\mathcal{S}_{m}$ can be recursively computed as follows:
\begin{align}\label{III.B.5-3}
\mathsf{V}^{\star}\left(\mathrm{S}_{m}\right)
&= \sum_{a\in\mathcal{A}}\pi^{\star}\left(\mathrm{S}_{m},a\right)\left(\mathsf{R}\left(\mathrm{S}_{m},\mathsf{U}\left(\mathrm{S}_{m}|a\right)\right)+ \gamma\mathsf{V}^{\star}\left(\mathsf{U}\left(\mathrm{S}_{m}|a\right) \right)\right),
\end{align}
where $\pi^{\star}\left(\mathrm{S}_{m},a\right)$ is the probability of choosing action $a$ at the state $\mathrm{S}_{m}$ according to the optimal policy.
In Fig.~\ref{FigOMDP}, we depict the state-action diagram of the MDP defined above. 
In this figure, the state $\mathrm{S}_{n}$ is transited to the next state $\mathsf{U}\left(\mathrm{S}_{n}|a\right)$ when taking an action $a$.
Particularly, when $a=1$, the state $\mathrm{S}_{n}$ is transited to the state $\mathsf{U}\left(\mathrm{S}_{n}|1\right)$ by exploiting the transmitted symbol index $k_{n}$.
Based on the state transition and the optimal policy in \eqref{III.B.5-1}, the states are transited to the next states until the end of data subblock.

Characterizing the optimal policy of the above MDP faces two major challenges in practical communication systems. 
First, the state transition is unknown at the receiver due to the lack of information of the transmitted symbol vectors.
Second, the number of the states in this MDP exponentially increases with the length of $T_{\rm u}$ (see Fig. \ref{FigOMDP}). 
To circumvent these challenges, in the following section, we design a computationally-efficient algorithm to solve the MDP without perfect knowledge on the state transition and the reward function.

\section{Proposed Channel Estimator via Reinforcement Learning}
RL is a type of machine learning that can find the optimal policy of an MDP with unknown or partial information on an environment's dynamics \cite{Sutton}.
In this section, we propose an efficient RL algorithm to approximately but efficiently determine the optimal policy of the MDP in Sec.~\ref{Sec:MDP}.
We then present the semi-data-aided channel estimator that utilizes the proposed RL algorithm for optimizing the symbol vector selection. 
We also introduce an additional strategy to improve detection performance after the symbol vector selection in the semi-data-aided channel estimator.

\subsection{Proposed RL Algorithm}\label{Sec:RL}
The key idea of the proposed RL algorithm is to exploit the APP computed from data detection for approximately determining the optimal policy based on MCTS \cite{Sutton,Browne2012,Vodopivec2017}.
In the proposed algorithm, we particularly modify the original MCTS to make this approach applicable for the receiver in practical systems. 
In what follows, we elaborate on the details of the proposed RL algorithm applied to determine the optimal policy for the state ${\rm S}_n \in \mathcal{S}_n$ with $n\in\{1,\ldots,T_{\rm u}\}$.

\subsubsection{Tree Policy and Rollout Policy}
The basic idea of MCTS is to determine the best action at the current state by looking ahead the rewards of near-future actions according to a {\em tree policy}, while approximating the rewards of distant-future actions according to a {\em rollout policy} \cite{Sutton}.
Typically, the tree policy is designed to mimic the optimal policy, while the design of the rollout policy focuses more on computational simplicity and tractability.  
To design an effective tree policy for the proposed algorithm, our intuition is that the higher the reliability of the detected symbol vector is, the higher the probability of selecting the corresponding symbol vector as an additional pilot signal is.
Inspired by this intuition, we exploit the APP computed from data detection as a measure of the reliability of the detected symbol vector.
We then set the tree policy of the proposed algorithm as
\begin{align}\label{eq:tree_pro}
{\pi}^{\rm t}\left(\mathrm{S}_{m},a\right)
&=\begin{cases} \hat{\theta}_{\hat{k}_{m}}[m], &a=1, \\ 1-\hat{\theta}_{\hat{k}_{m}}[m], &  a=0, \end{cases}
\end{align}
for every state $\mathrm{S}_{m}\in \mathcal{S}_{m}$ with $m\in\{n+1,\ldots,n+N\}$, where $N$ is the number of the near-future actions taken according to the tree policy. 
We also denote the sequence of actions randomly chosen by the tree policy in \eqref{eq:tree_pro} by ${\bf a}^{\rm t} = [{a}_{1}^{\rm t},\cdots,a_{N}^{\rm t}] \in\mathcal{A}^{N}$.
To determine an effective rollout policy, we introduce a pre-determined threshold $\eta_{\rm roll}$ such that $0 \leq \eta_{\rm roll} \leq 1$. 
We then choose the action $a=1$ if the APP is higher than $\eta_{\rm roll}$ and $a=0$ otherwise, i.e.,
\begin{align}\label{eq:rollout_pro}
    {\pi}^{\rm r}\left(\mathrm{S}_{m}\right)
    &=\begin{cases} 1, & \hat{\theta}_{\hat{k}_{m}}[m] \geq \eta_{\rm roll}, \\ 0, &  \hat{\theta}_{\hat{k}_{m}}[m] < \eta_{\rm roll}, \end{cases}
\end{align}
for every state $\mathrm{S}_{m}\in \mathcal{S}_{m}$ associated with time slot $m \in\{n+N+1,\ldots,T_{\rm u}\}$.
Our rollout policy is useful to reduce the computational complexity of the value function estimation after $N$ state transitions.
Meanwhile, this policy also mimics the behavior of the tree policy \eqref{eq:tree_pro} if the detected symbol vector is reliable enough (i.e., $\theta_{\hat{k}_{m}}[m] \approx 1$). 
We denote the sequence of actions determined by the rollout policy in \eqref{eq:rollout_pro} by ${\bf a}^{\rm r} = [{a}_1^{\rm r},\cdots,a_{T_{\rm u}-n-N}^{\rm r}] \in\mathcal{A}^{T_{\rm u}-n-N}$.

\begin{figure*}
	\centering
	\epsfig{file=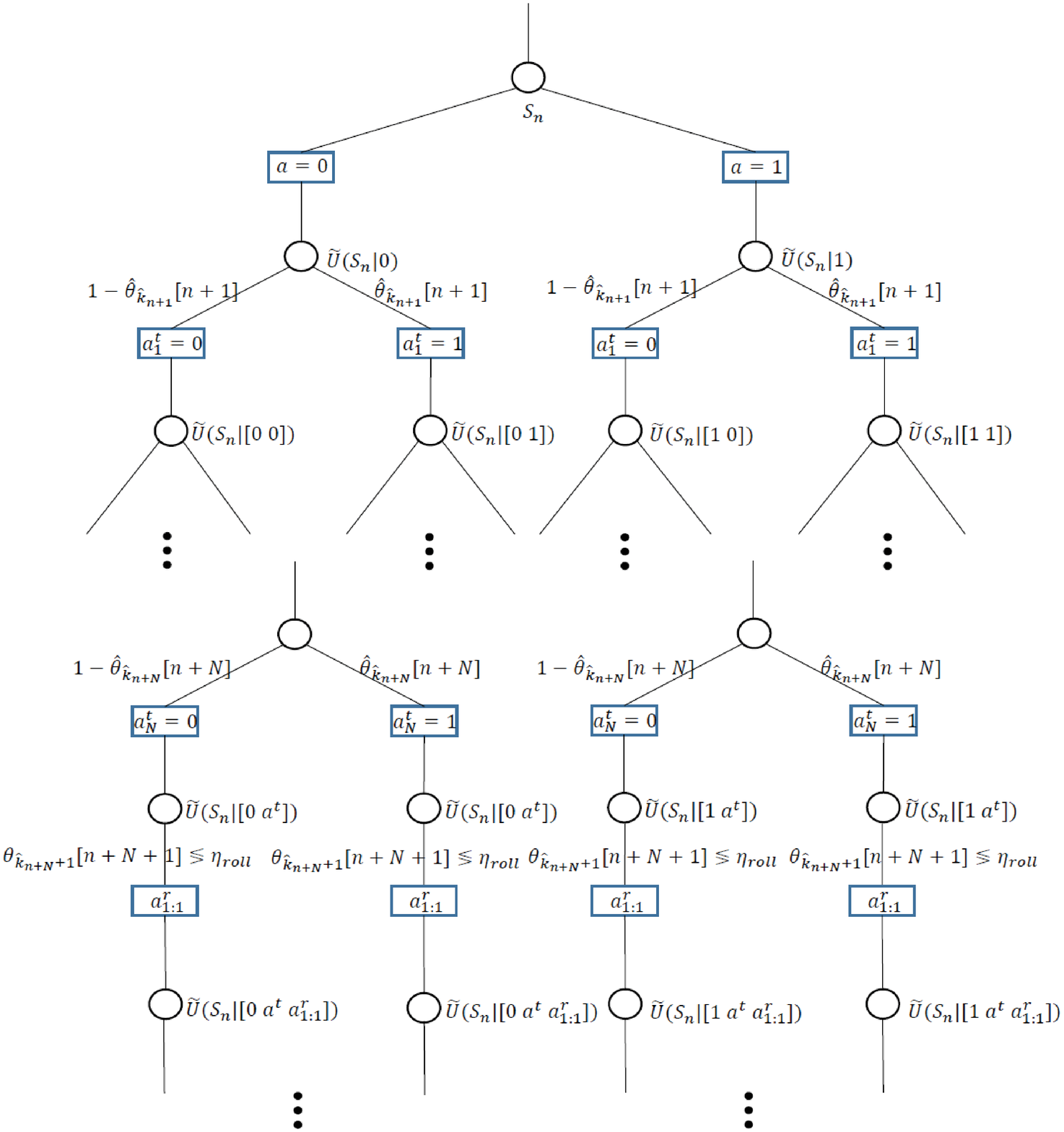, width=16cm}
	\caption{State-action diagram of the approximate MCTS for $a\in\mathcal{A}$ and $\mathrm{S}_{n}\in\mathcal{S}_{n}$.}
	\label{FigMDP}
\end{figure*}

\subsubsection{Approximation for Monte Carlo Simulations}
In the original MCTS, the optimal value function is estimated through Monte Carlo simulations according to the tree policy and the rollout policy \cite{Sutton}.
Unfortunately, a receiver in practical communication systems cannot adopt such simulation-based approach because executing the Monte Carlo simulations requires perfect information of the transmitted symbol vectors at the receiver.
To circumvent this limitation, we introduce a virtual state that can mimic the effect of Monte Carlo simulations without actual execution.
The virtual state is defined as the state that can be arrived when the true symbol vector exactly behaves like its expectation:
\begin{align}\label{soft_decision}
    \mathbb{E} \big[{\bf x}[m] \big| {\bf y}[m], {\bf H} \big]
    = \sum\limits_{j=1}^{K} \theta_{j}[m] \mathbf{x}_{j},
\end{align}
for $m\in\{n,\ldots,T_{\rm u}\}$.
We refer to the expectation in \eqref{soft_decision} as an \textit{expected} symbol vector at time slot $m$. 
Since the receiver cannot compute the exact APP due to the lack of perfect channel knowledge, we use an approximate APP for tracking both the tree policy and the rollout policy.
When tracking the tree policy, we use an accurate estimate based on a new channel estimate obtained by taking the series of actions according to the tree policy. 
Let ${\bf a}^{\rm t} \in \mathcal{A}^N$ be the sequence of the actions chosen by the tree policy in  \eqref{eq:tree_pro}.
When taking the actions in ${\bf a}^{\rm t}$ from the state ${\rm S}_n= \big({\bf X}_n,\hat{\bf X}_n,{\bf a}_n\big) \in \mathcal{S}_n$, the LMMSE channel estimate is given by 
\begin{align}\label{channel_tree}
    \hat{\bf H}\left(\mathrm{S}_{n};{\bf a}^{\rm t}\right) 
    &= \bar{\bf Y}\left(\mathrm{S}_{n};{\bf a}^{\rm t}\right)\bar{\bf X}^H (\mathrm{S}_{n};{\bf a}^{\rm t}) \big(\bar{\bf X}(\mathrm{S}_{n};{\bf a}^{\rm t})  \bar{\bf X}^H(\mathrm{S}_{n};{\bf a}^{\rm t}) +\sigma^{2}\mathbf{I}_{N_{\rm{tx}}}\big)^{-1},
\end{align}
where 
\begin{align*}
    \bar{\bf X}(\mathrm{S}_{n};{\bf a}^{\rm t}) &= \big[\hat{\bf X}_n, \mathbf{\tilde{x}}[n+l_1({\bf a}^{\rm t})],\cdots,\mathbf{\tilde{x}}[n+l_{\|{\bf a}^{\rm t}\|_0}({\bf a}^{\rm t})] \big], \\
    \bar{\bf Y}\left(\mathrm{S}_{n};{\bf a}^{\rm t}\right)&=\big[{\bf Y}({\rm S}_n),{\bf y}[n+l_1({\bf a}^{\rm t})]\cdots,{\bf y}[n+l_{\|{\bf a}^{\rm t}\|_0}({\bf a}^{\rm t})]  \big].
\end{align*}        
Based on the channel estimate in \eqref{channel_tree}, the APP estimate used for tracking the tree policy is determined as
\begin{align}
    \hat{\theta}_{k}^{\rm t}[m] = \frac{ \exp\big(-\frac{1}{\sigma^{2}}\|\mathbf{y}[m]-\hat{\bf H}\big(\mathrm{S}_{n};{\bf a}^{\rm t}\big)\mathbf{x}_{k}\|^{2}\big)}
    {\sum_{j\in\mathcal{K}}\exp\big(-\frac{1}{\sigma^{2}}\|\mathbf{y}[m]-\hat{\bf H}\big(\mathrm{S}_{n};{\bf a}^{\rm t}\big)\mathbf{x}_{j}\|^{2}\big)}.
\end{align}
Unlike the tree policy, we use the initial estimate of the APP in \eqref{eq:APP_approx} when tracking the rollout policy, in order to reduce a required computational complexity. 
Utilizing the above strategy, we approximate the expected symbol vector as
\begin{align}\label{soft_decision2}
    \mathbf{\tilde{x}}[m] 
    = \begin{cases}
    \sum\limits_{j=1}^{K} \hat{\theta}_{j}^{\rm t}[m] \mathbf{x}_{j}, &  m\in\{n,n+1,\ldots,n+N\}, \\
    \sum\limits_{j=1}^{K} \hat{\theta}_{j}[m] \mathbf{x}_{j}, &  m\in\{n+N+1,\ldots, T_{\rm u}\}.
    \end{cases}
\end{align}
Under the assumption of ${\bf x}[m] = \tilde{\bf x}[m]$ for $m\in\{n,\ldots,T_{\rm u}\}$, the virtual state arrived when taking the sequence of actions ${\bf a}=[a_1,\ldots,a_m] \in \mathcal{A}^m$ from the state $\mathrm{S}_{n} \in \mathcal{S}_{n}$ is given by
\begin{align}\label{III.C.2-4}
	\mathsf{\tilde{U}}\left(\mathrm{S}_{n}|{\bf a}\right)
	&= \left(\big[{\bf X}_{n},\tilde{\bf X}_n({\bf a})\big], 
	\big[\hat{\bf X}_{n},\hat{\bf X}_n({\bf a})\big],
	[{\bf a}_n, {\bf a}]\right),
\end{align}
where 
\begin{align}
    \tilde{\bf X}_n({\bf a}) &= \big[\mathbf{\tilde{x}}[n+l_1({\bf a})],\cdots,\mathbf{\tilde{x}}[n+l_{\|{\bf a}\|_0}({\bf a})] \big],
    \nonumber\\
    \hat{\bf X}_n({\bf a}) &= \big[\mathbf{\hat{x}}[n+l_1({\bf a})],\cdots,\mathbf{\hat{x}}[n+l_{\|{\bf a}\|_0}({\bf a})] \big].\nonumber
\end{align}
By using our policies \eqref{eq:tree_pro}, \eqref{eq:rollout_pro} and virtual state in \eqref{III.C.2-4}, we depict the state-action diagram of our MCTS approach in Fig.~\ref{FigMDP}. 
The tree policy that mimics the optimal policy is applied for the time indices $\{n+1,\ldots,n+N\}$.
Because the tree policy considers the effect of both actions, the number of states to compute is proportional to $2^{N}$.
In contrast, the rollout policy only considers the reliable state which has a higher APP and therefore requires a much lower computational complexity.

\subsubsection{Optimal Policy}
Based on the MCTS approach explained above, we characterize the optimal policy of the MDP in Sec.~\ref{Sec:MDP} in a closed-form expression.
This result is given in the following theorem:
\begin{thm}
When employing the MCTS approach in Sec.~\ref{Sec:RL}, the optimal policy of the MDP in Sec.~\ref{Sec:MDP} for a state $\mathrm{S}_{n} =\big(\mathbf{X}_{n},\mathbf{\hat{X}}_{n},\mathbf{a}_{n} \big)\in\mathcal{S}_{n}$ is 
\begin{align}\label{III.C.3-5}
    \pi^{\star}(\mathrm{S}_{n}) 
    &=\mathbb{I}\Bigg[\sum_{\mathbf{a}^{\rm{t}}\in\mathcal{A}^{N}} \omega_n^{\rm t}(\mathbf{a}^{\rm{t}})
    \Delta_n([\mathbf{a}^{\rm{t}},\mathbf{a}^{\rm{r}}])\geq 0\Bigg],
\end{align}
where
\begin{align}
\omega_n^{\rm t}(\mathbf{a}) &=  \prod\limits_{l=1}^{|{\bf a}|}~ (\hat{\theta}_{\hat{k}_{n+l}}[n+l])^{a_{l}}\big(1-\hat{\theta}_{\hat{k}_{n+l}}[n+l]\big)^{1-a_{l}}, \label{eq:weight} \\
\Delta_n(\mathbf{a}) & = \|\mathbf{{t}}_{n}({\bf a})\|^{2} \Big\{ \sigma^{2} +\sigma^{4} \big( \|\mathbf{{t}}_{n}({\bf a})\|^{2} - 2{\beta}_{n}({\bf a}) \big)  +\|\mathbf{{v}}_{n}({\bf a})\|^{2} - \|\mathbf{{e}}_{n}({\bf a})-\mathbf{{u}}_{n}({\bf a})+\mathbf{{v}}_{n}({\bf a})\|^{2} \Big\},  \label{eq:Delta_n}
\end{align}
and related parameters are defined as
\begin{align}
{\bf Q}_n({\bf a})&=\left(\mathbf{\hat{X}}_{n}\mathbf{\hat{X}}_{n}^{H}+\hat{\bf X}_n({\bf a}) \hat{\bf X}_n^{H}({\bf a})+\sigma^{2}\mathbf{I}_{N_{\rm{tx}}}\right)^{-1},\nonumber\\
{\bf D}_n({\bf a})&=\mathbf{\hat{X}}_{n}(\mathbf{\hat{X}}_{n}-\mathbf{X}_{n})^{H} + \hat{\bf X}_n({\bf a}) (\hat{\bf X}_n({\bf a}) - \tilde{\bf X}_n({\bf a}))^{H}+\sigma^{2}\mathbf{I}_{N_{\rm{rx}}} , \nonumber \\
{\bf t}_{n}({\bf a})&=\frac{1}{\sqrt{1+\alpha_n({\bf a})}}{\bf Q}_{n}({\bf a})\mathbf{\hat{x}}[n],\nonumber\\ 
{\bf e}_n({\bf a}) &= \frac{1}{\sqrt{1+\alpha_n({\bf a})}}( \mathbf{\hat{x}}[n]-\mathbf{\tilde{x}}[n]),\nonumber\\
\mathbf{{u}}_{n}({\bf a})&={\bf D}_n^H({\bf a}) {\bf t}_{n}({\bf a}),\nonumber\\
\mathbf{{v}}_{n}({\bf a})&=\frac{1}{\|{\bf t}_{n}({\bf a})\|^2}{\bf D}_n^H({\bf a}) {\bf Q}_{n}({\bf a}){\bf t}_{n}({\bf a}),\nonumber\\
{\alpha}_{n}({\bf a})&=\mathbf{\hat{x}}^{H}[n]{\bf Q}_{n}({\bf a})\mathbf{\hat{x}}[n], \nonumber\\
{\beta}_{n}({\bf a})&=\frac{1}{\|{\bf t}_{n}({\bf a})\|^2}{\bf t}_{n}^H({\bf a}){\bf Q}_{n}({\bf a}){\bf t}_{n}({\bf a}). \nonumber
\end{align}

\end{thm}
\begin{IEEEproof}	
	See Appendix A.	
\end{IEEEproof}
The optimal policy in \eqref{III.C.3-5} determines the best action at the current state by considering the average reward of all possible $N$ future actions that can be chosen by the tree policy. 
In this context, the weight $\omega_n^{\rm t}(\mathbf{a}^{\rm t})$ in \eqref{eq:weight} represents the probability of taking a certain sequence of actions, ${\bf a}^{\rm t}$, according to the tree policy in \eqref{eq:tree_pro}.
As can be seen from Theorem~1, the most prominent feature of the proposed RL algorithm is that the optimal policy has a closed-form expression which can be computed in a deterministic way at the receiver in practical systems.

\subsubsection{Low-Complexity Policy}
A major limitation of the optimal policy in \eqref{III.C.3-5} is that the complexity of computing the policy increases exponentially with the number of the near-future actions, $N$.
Therefore, computing this policy is not affordable if $N$ is large, implying that we cannot arbitrarily increase the value of $N$ to improve the performance of the proposed RL algorithm. 
To circumvent this limitation, we develop a low-complexity policy that approximates the optimal policy in \eqref{III.C.3-5} based on Monte Carlo sampling. 
Recall that the weighted sum in \eqref{III.C.3-5} is nothing but the expectation of $\Delta_n([\mathbf{a}^{\rm{t}},\mathbf{a}^{\rm{r}}])$ because $\omega_n^{\rm t}(\mathbf{a}^{\rm t})$ is the probability of obtaining an action sequence ${\bf a}^{\rm t}$ from the tree policy in \eqref{eq:tree_pro}. 
Motivated by this observation, we randomly draw $N_{\rm sample}$ samples of ${\bf a}^{\rm t}$ according to the tree policy. 
We then compute the {\em empirical} mean of $\Delta_n([\mathbf{a}^{\rm{t}},\mathbf{a}^{\rm{r}}])$ by averaging the values of $\Delta_n([\mathbf{a}^{\rm{t}},\mathbf{a}^{\rm{r}}])$ computed for $N_{\rm s}$ samples. 
Let $\hat{\Delta}_n$ be the empirical mean determined by the Monte Carlo sampling approach. 
Then the optimal policy is approximately determined as $\pi^{\star}(\mathrm{S}_{n}) = \mathbb{I}[\hat{\Delta}_n \geq 0]$.
We refer to this policy as a {\em low-complexity} policy for the state ${\rm S}_n$. 
The complexity required for determining the low-complexity policy increases linearly with the number of samples, $N_{\rm sample}$, and is independent from $N$. 
Therefore, the overall complexity of the proposed RL algorithm can be significantly reduced by harnessing the low-complexity policy with $N_{\rm sample} \ll 2^N$. 
It is also possible to reduce a mismatch between the optimal policy and the low-complexity policy by increasing $N_{\rm sample}$ at the cost of the complexity.

\vspace{3mm}
{\bf Remark (Applicability to other data detection methods):} 
A key requirement of the proposed RL algorithm is the APPs that can be directly obtained from the MAP data detection method. 
Despite this requirement, the proposed RL algorithm is universally applicable to any other soft-output data detection method which computes the log-likelihood ratios (LLRs) of transmitted data bits.
In this case, the proposed RL algorithm can utilize the APPs computed from the LLRs which can be readily performed at the receiver with a slight increase in the computational complexity.  

\subsection{Proposed Channel Estimator}
The proposed channel estimator adopts the RL algorithm in Sec.~\ref{Sec:RL} for optimizing the selection of detected symbol vectors utilized as additional pilot signals. 
The proposed channel estimator is summarized in Algorithm 1.
In particular, depending on the choice of a policy determination strategy, the receiver computes either the optimal policy in Step $5$ or the low-complexity policy in Steps $7$--$12$. 
In Step $14$, we consider the most probable state transition for the unknown transmitted symbol vector.
To address this, the detected symbol vector $\hat{k}_{n}$ is assumed to be the same as the transmitted symbol vector if the action $1$ is chosen according to the optimal policy.
The corresponding state $\mathsf{\hat{U}}\left(\mathrm{S}_{n}|a\right)\in\mathcal{S}_{n+1}$ is given by,
\begin{align}\label{IV.B.1-1}
\mathsf{\hat{U}}\left(\mathrm{S}_{n}|a\right)
&=\begin{cases} \big([\mathbf{X}_{n},\mathbf{x}_{\hat{k}_{n}}],[\mathbf{\hat{X}}_{n},\mathbf{\hat{x}}[n]],[{\bf a}_n,1]\big),& a=1, \\
\big(\mathbf{X}_{n},\mathbf{\hat{X}}_{n},[{\bf a}_n,0]\big),&a=0. \end{cases}
\end{align}
Finally, at time slot $T_{\rm u}$, we can obtain the updated channel estimate $\mathbf{\hat{H}}^{u} = \mathbf{\hat{H}}(\mathrm{S}_{T_{\rm u}+1})$.

\begin{algorithm}[ht]
    {\small 
	\SetAlgoNoLine%
	\DontPrintSemicolon
	\caption{The proposed semi-data-aided channel estimator}
	{
		Set $\mathrm{S}_{1}=\left({\bf P},{\bf P},\phi\right)$.\\
	}
    {
		\For{$n=1$~{\rm to}~$T_{\rm u}$}
		    {
		    Determine ${\bf a}^{\rm r}$ according to \eqref{eq:rollout_pro}.\\
		    \uIf{Optimal policy}{
                Compute $a^{\star} = \pi^{\star}(\mathrm{S}_{n})$ from \eqref{III.C.3-5}.\\
            }
            \ElseIf{Low-complexity policy}{
                Initialize $\hat{\Delta}_n = 0$.\\
    	        \For{$s=1$~{\rm to}~$N_{\rm sample}$}{
    	        Randomly draw ${\bf a}^{\rm t}$ according to \eqref{eq:tree_pro}. \\ 
                Update $\hat{\Delta}_n \leftarrow \frac{s-1}{s}\hat{\Delta}_n  +\frac{1}{s} \Delta_n\left([\mathbf{a}^{\rm{t}},\mathbf{a}^{\rm{r}}]\right)$ from \eqref{eq:Delta_n}. \\ 
                }
                Set $a^{\star} = \pi^{\star}(\mathrm{S}_{n}) = \mathbb{I}[\hat{\Delta}_n \geq 0]$.\\
		    }
            Set $\mathrm{S}_{n+1} = \mathsf{\hat{U}}\left(\mathrm{S}_{n}|a^\star\right)$ from \eqref{IV.B.1-1}.\\
		    }
		Set $\mathbf{\hat{H}}^{u} = \mathbf{\hat{H}}(\mathrm{S}_{T_{\rm u}+1})$ from \eqref{III.B.1-2}.
	}}
\end{algorithm}

\begin{figure}
	\centering
	\epsfig{file=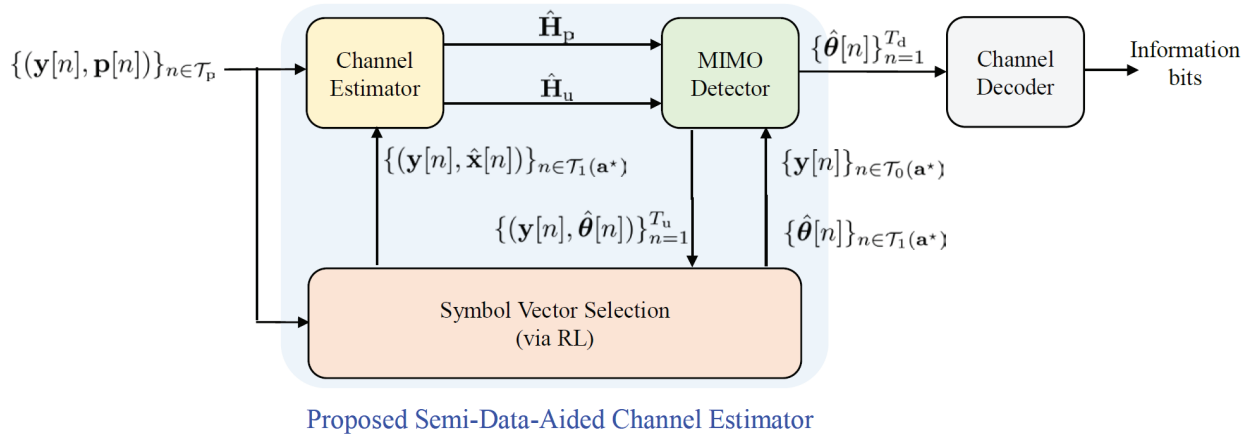, width=16.7cm}
	\caption{A block diagram of receive processing with the proposed semi-data-aided channel estimator.}
	\label{Fig:Proposed}
\end{figure}

\subsection{Re-Detection of Unselected Symbol Vectors}
An important byproduct of the proposed semi-data-aided channel estimator is the set of detected symbol vectors that are not selected as pilot signals for channel estimation. 
Since these vectors are turned out to be not {\em reliable}, we can treat them as incorrectly detected symbol vectors. 
Motivated by this observation, to further improve detection performance, we utilize the final channel estimate determined by Algorithm~1 for re-detecting received signals associated with the symbol vectors not selected by the proposed RL algorithm. 
Suppose that the final state and the channel estimate of Algorithm~1 is given by ${\rm S}^\star = \big({\bf X}^\star,\hat{\bf X}^\star, {\bf a}^\star\big)$ and $\hat{\bf H}^{\star}$, respectively.
Then the set of time slot indices associated with the unselected symbol vectors is expressed as
\begin{align}
    \mathcal{T}_{0}({\bf a}^\star) = \{l ~|~ a_l^\star = 0 \}, 
\end{align}
where $a_l^\star$ is the $l$-th entry of ${\bf a}^\star$. 
The optimal MAP data detection is performed again based on $\hat{\bf H}^{\star}$ for the received signals associated with time slots in $\mathcal{T}_{0}({\bf a}^\star)$.
This strategy yields 
\begin{align}
    \hat{k}_n = \argmax\limits_{k\in\mathcal{K}}~\hat{\theta}_{k}^\star [n],~~\forall n \in \mathcal{T}_{0}({\bf a}^\star),
\end{align}
where 
\begin{align}
    \hat{\theta}_{k}^\star[n] = \frac{ \exp\big(-\frac{1}{\sigma^{2}}\|\mathbf{y}[n]-\hat{\bf H}^{\star}\mathbf{x}_{k}\|^{2}\big)}
    {\sum_{j\in\mathcal{K}}\exp\big(-\frac{1}{\sigma^{2}}\|\mathbf{y}[n]-\hat{\bf H}^{\star}\mathbf{x}_{j}\|^{2}\big)}.
\end{align}

In Fig.~\ref{Fig:Proposed}, we illustrate the overall receive processing with the proposed semi-data-aided channel estimator and the re-detection strategy, where $\mathcal{T}_{1}({\bf a}^\star) = \{l ~|~ a_l^\star = 1\}$ and $\hat{\bm \theta}[n] = \big[\hat{\theta}_1[n],\cdots, \hat{\theta}_K[n]\big]^{T}$.
Although the re-detection process requires an additional complexity, this process is executed only once more for a portion of received signals. 
Therefore, the complexity of our strategy is still lower than that of iterative data-aided channel estimation (e.g., \cite{Kim2012, Verenzuela2020, Ma2014, Huang2018, Zhao2008, Park2015}) which requires multiple executions of channel estimation and data detection for the whole received signals.

\section{Simulation Results}

In this section, using simulations, we evaluate the NMSE and BLER of the proposed channel estimator in a coded MIMO system with the MAP data detection method.
In these simulations, we consider 4-QAM for symbol mapping and assume that  $N_{\rm{tx}}=2$, $N_{\rm{rx}}=4$, $T_{\rm{p}}=4$, $T_{\rm{u}}=200$, and $T_{\rm{d}} = 2048$.
For channel coding, we adopt the rate $1/2$ turbo code based on parallel concatenated codes with feedforward and feedback polynomial $(15,13)$ in octal notation.
For performance comparison, we consider the following methods:
\begin{itemize}
    \item {\bf PCSI:} This is an ideal case in which perfect channel state information is available at the receiver (i.e., $\hat{\bf H} = {\bf H}$). 
    
    \item {\bf Pilot-CE:} This is a conventional pilot-aided channel estimator described in Sec. II-B.
    
    \item {\bf Semi-CE (Opt):} This is a semi-data-aided channel estimator when correctly detected symbol vectors are utilized as additional pilot signals by assuming perfect knowledge of transmitted symbol vectors. This can be interpreted as the {\em true} optimal policy of the MDP in Sec. III-C. 
    
    \item {\bf Semi-CE (Pro, Opt):} This is a semi-data-aided channel estimator when detected symbol vectors are selected according to the optimal policy determined by the proposed RL algorithm. A re-detection strategy discussed in Sec.~IV-C is also adopted. 
    
    \item {\bf Semi-CE (Pro, Low):} This is a semi-data-aided channel estimator when detected symbol vectors are selected according to the low-complexity policy determined by the proposed RL algorithm. A re-detection strategy discussed in Sec.~IV-C is also adopted. 
    
    \item {\bf Semi-CE (All):} This is a semi-data-aided channel estimator when all the expected symbol vectors in \eqref{soft_decision} are utilized as additional pilot signals.
    
    \item {\bf Iter-CE:} This is an iterative data-aided channel estimator in which the best $T_{u}$ virtual pilots are utilized as additional plot signals for every iteration. The number of iterations is set as $4$. This method is a slight modification of the method proposed in \cite{Park2015}.
\end{itemize}
We set the parameters of the proposed RL algorithm as $(N,N_{\rm sample},\eta_{\rm{roll}})=(8,10,0.5)$ unless specified otherwise.
We define a per-bit signal-to-noise ratio (SNR) as $E_{b}/N_{0} = \frac{1}{\log_{2} |\mathcal{X}|\sigma^{2}}$, and also define NMSE as $\frac{\|\hat{\bf H} - {\bf H} \|_{\rm F}^{2} }{ \|{\bf H}\|_{\rm F}^{2}}$.


\begin{figure}
	\centering
	\subfigure[NMSE performance]{\epsfig{file=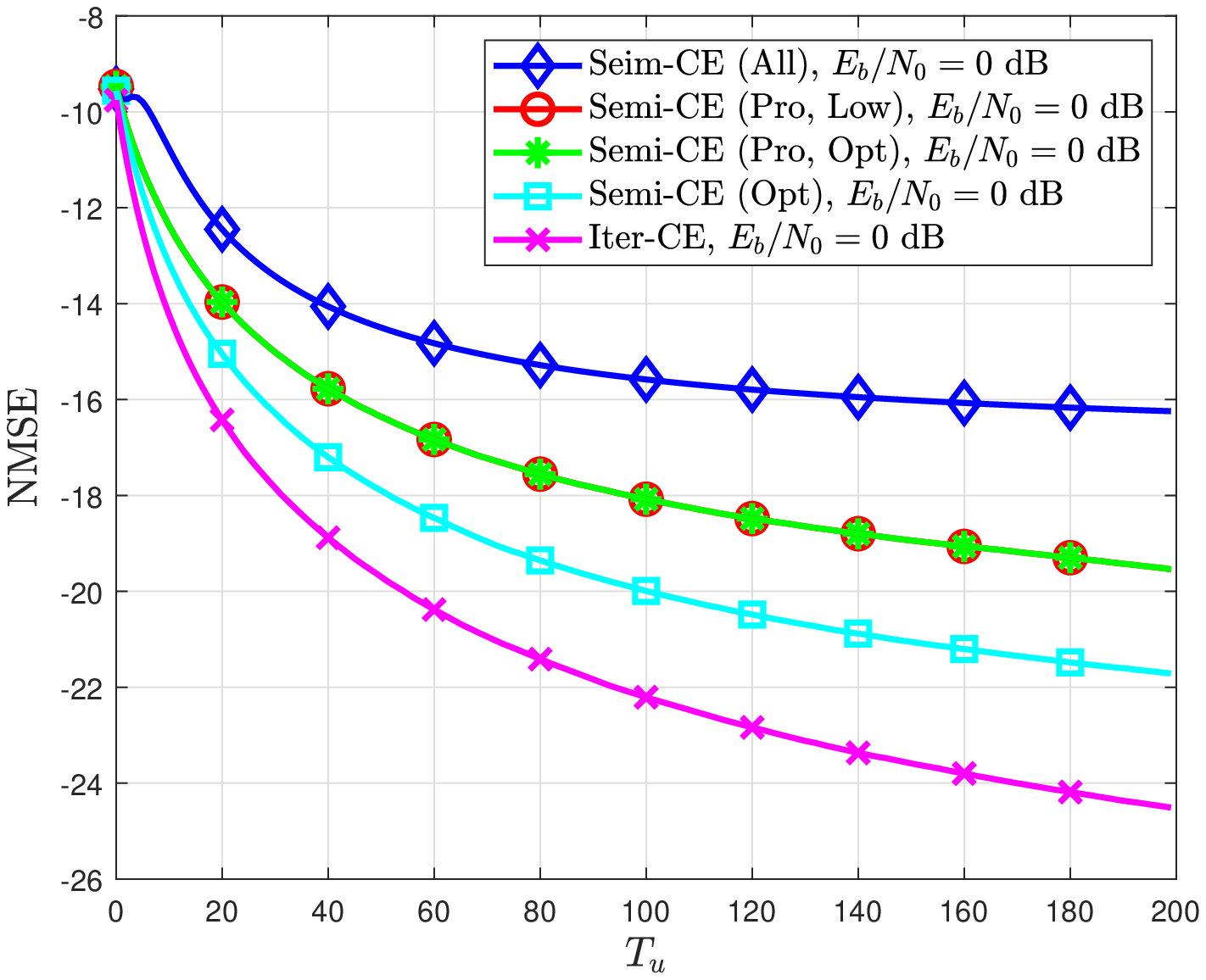, width=8.1cm}}
	\subfigure[BLER performance]{\epsfig{file=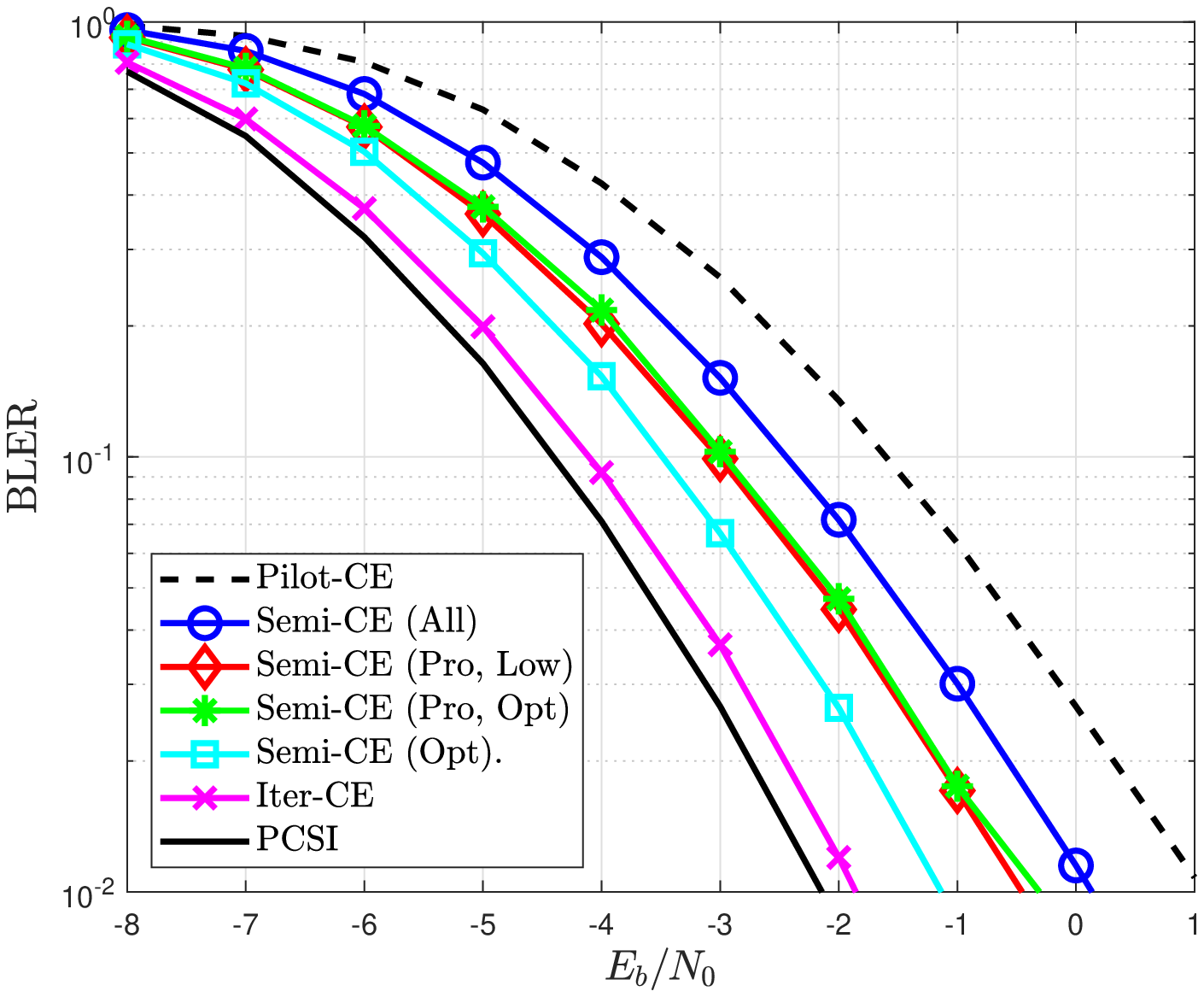, width=8.1cm}}
	\caption{NMSE and BLER of various channel estimators for different per-bit SNRs.}
	\label{FigDet}
\end{figure}

Fig. \ref{FigDet} compares the NMSE and BLER of various channel estimators for different per-bit SNRs.
Fig. \ref{FigDet} shows the proposed channel estimator has better NMSE and BLER performances than the conventional pilot-aided channel estimator by exploiting detected symbol vectors as additional pilot signals. 
It is also shown that the proposed channel estimator outperforms {\bf Semi-CE (All)} which exploits all the detected symbol vectors without a proper selection. 
Meanwhile, the SNR gap between the proposed channel estimator and {\bf Semi-CE (Opt)} is only $0.5$ dB.
These results demonstrate that the proposed RL algorithm effectively selects a set of detected symbol vectors that can improve the performance of the semi-data-aided channel estimation.
Another interesting observation is that {\bf Semi-CE (Pro, Low)} has a similar performance to {\bf Semi-CE (Pro, Opt)}; this result implies that the low-complexity policy, whose complexity is proportional to $N_{\rm sample}=10$, well approximates the optimal policy, whose complexity is proportional to $2^{N=8}=256$, by leveraging a Monte Carlo sampling method. 
Although {\bf Iter-CE} achieves the best performance among the considered channel estimators, it significantly increases both the delay and computational complexity of the overall receive processing because this estimator requires repeated executions of both data detection and channel decoding. 

\begin{figure}
	\centering
	\subfigure[NMSE performance]{\epsfig{file=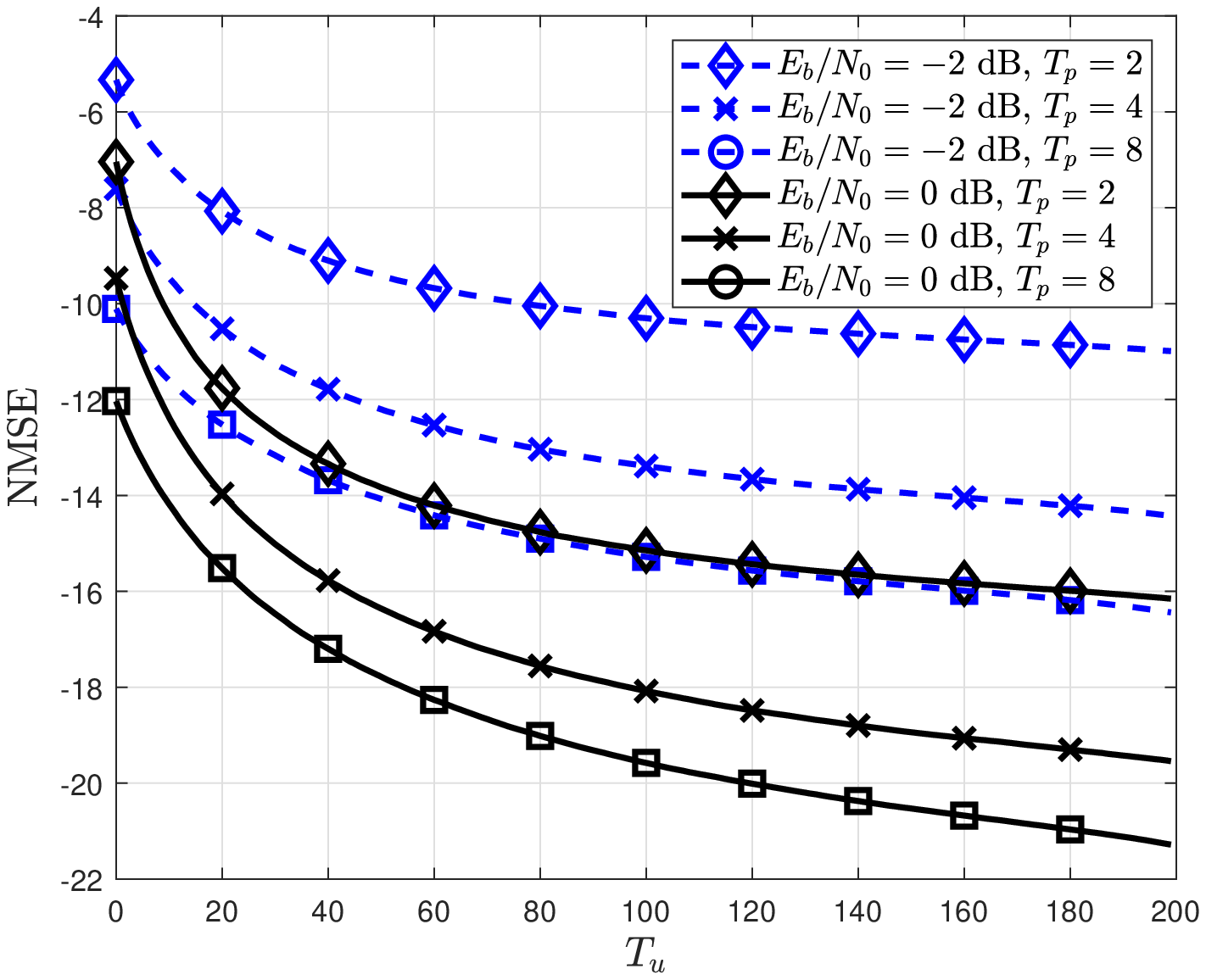, width=8.1cm}}
	\subfigure[BLER performance]{\epsfig{file=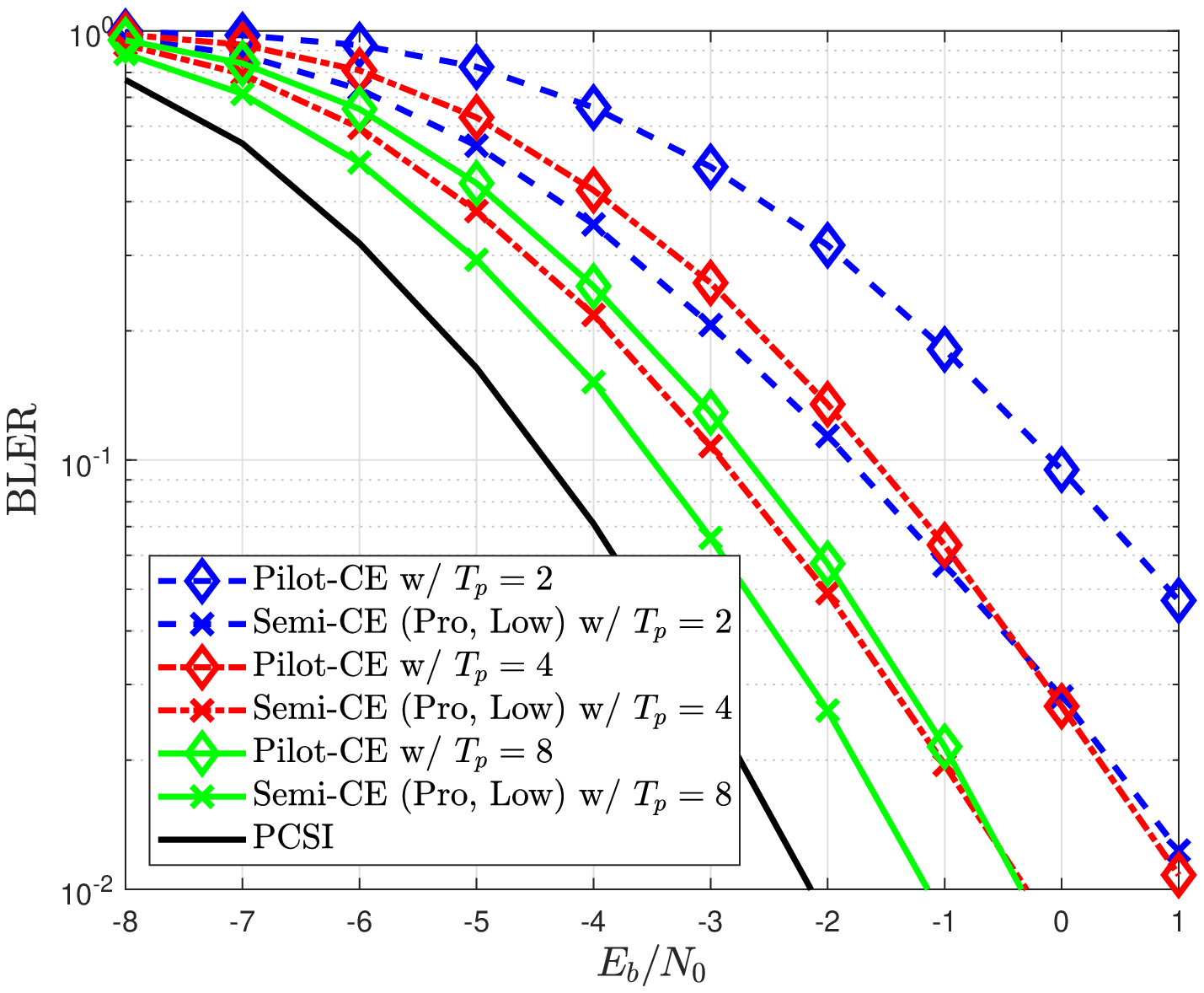, width=8.1cm}}
	\caption{Performance comparison of various channel estimators for different pilot lengths.}
	\label{FigTp}
\end{figure}

Fig. \ref{FigTp} compares the NMSE and BLER of various channel estimators for different pilot lengths.
Fig. \ref{FigTp} shows that the proposed channel estimator provides significant performance gain compared to the conventional pilot-aided channel estimator regardless of the pilot length.
It is also shown that a larger NMSE reduction is achieved in the case of $E_{b}/N_{0}=-2$ dB than in the case of $E_{b}/N_{0}=0$ dB.
The reason behind this result is that the number of reliable detected symbol vectors increases as the detection performance improves, which allows the use of a more accurate MCTS approach in the proposed RL algorithm.
Another interesting observation in Fig. \ref{FigTp}(b) is that the proposed channel estimator with $T_{\rm{p}} = 4$ even outperforms {\bf Pilot-CE} with $T_{\rm{p}} = 8$, which implies that the proposed estimator requires fewer pilot signals to achieve the same BLER performance.

\begin{figure}
	\centering
	\subfigure[NMSE performance]{\epsfig{file=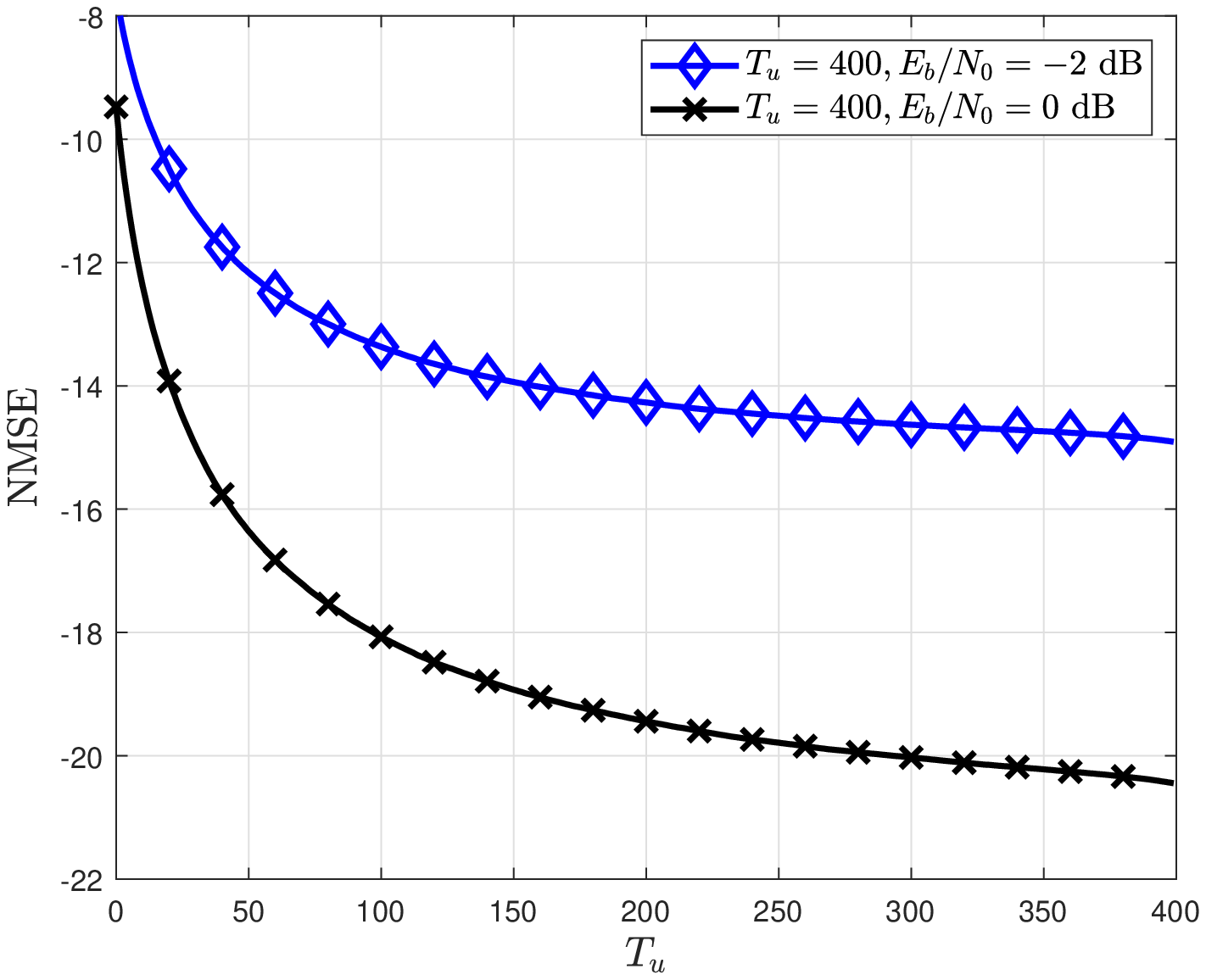, width=8.1cm}}
	\subfigure[BLER performance]{\epsfig{file=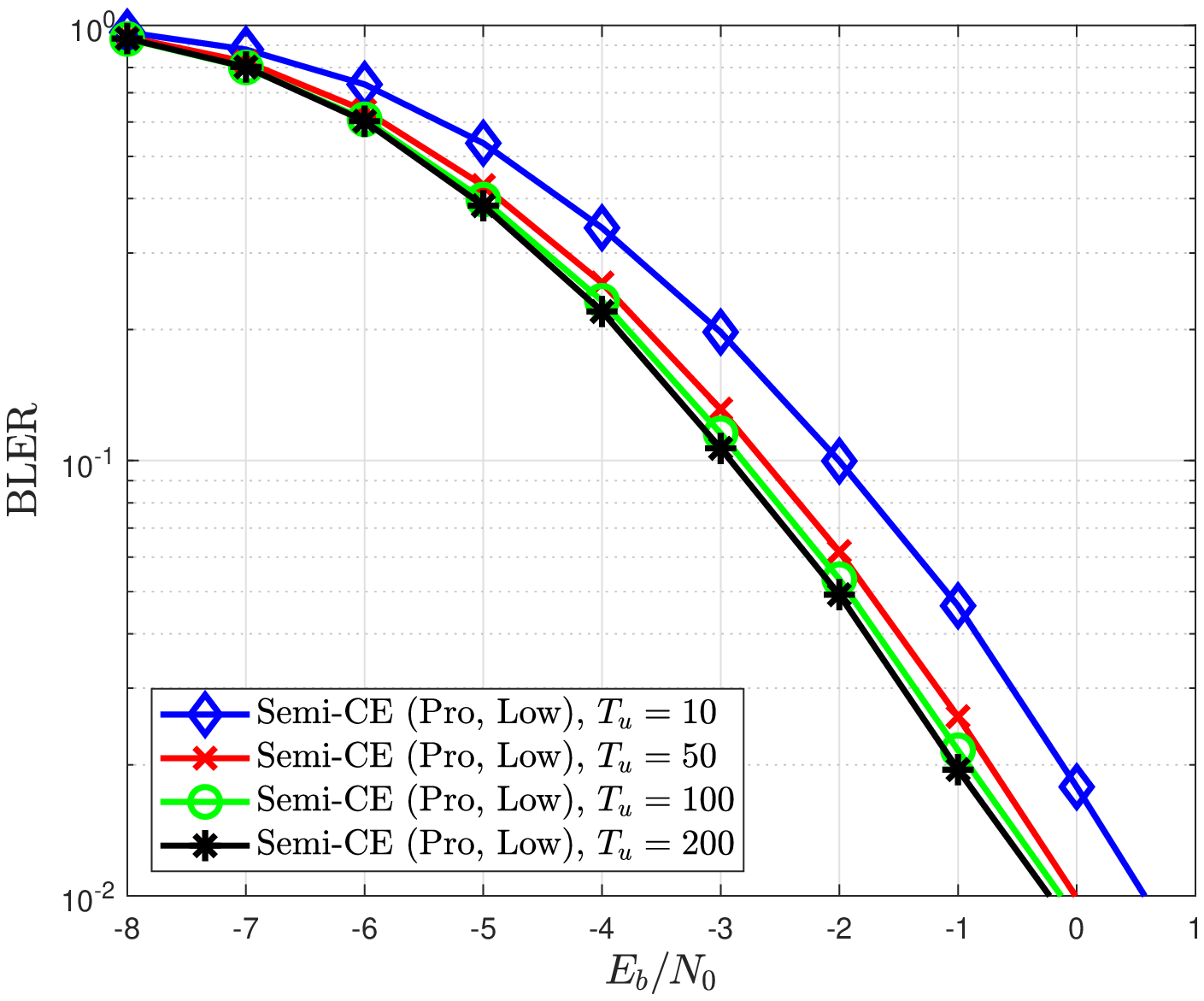, width=8.1cm}}
	\caption{Performance comparison of various channel estimators for different $T_{\rm{u}}$.}
	\label{FigTu}
\end{figure}

\begin{figure}
	\centering
	\epsfig{file=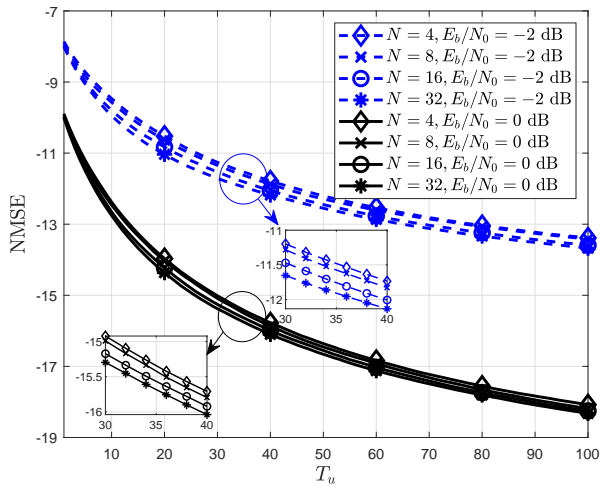, width=9cm}
    \caption{NMSE performance of the proposed channel estimator according the number of near-future actions $N$.}
	\label{FigTI}
\end{figure}

Fig. \ref{FigTu} compares the NMSE and BLER of the proposed channel estimator for different $T_{\rm{u}}$. 
Fig. \ref{FigTp}(a) shows that the NMSE performance of the proposed channel estimator improves with $T_{\rm{u}}$.
This gain is attained by increasing the number of detected symbol vectors that can be utilized as additional pilot signals.
Thanks to this gain, in Fig. \ref{FigTp}(b), it is shown that the BLER performance with the proposed channel estimator also improves with $T_{\rm u}$. 
Another important observation is that the improvement of both the NMSE and BLER performances decreases as $T_{\rm{u}}$ increases.
This result implies that once a sufficient number of additional pilot signals are attained, there is no significant gain by increasing the number of pilot signals, while the computational complexity of data-aided channel estimation is proportional to $T_{\rm u}$. 
Considering this fact, the semi-data-aided channel estimation is an effective strategy for adjusting the performance-complexity trade-off of data-aided channel estimation, by controlling the length of $T_{\rm u}$. 


Fig. \ref{FigTI} compares the NMSE of the proposed channel estimator for different numbers of the near-future actions, $N$, in the MCTS approach.
Fig. \ref{FigTI} shows that the NMSE performance of the proposed channel estimator improves with $N$ because increasing this number allows the proposed RL algorithm to accurately estimate  near-future rewards. 
This performance gain, however, is attained at the cost of the computational complexity required for determining the best policy for the MDP. 
Considering this trade-off, in our simulations, we set $N=8$ which provides sufficient accuracy for the estimation of the near-future rewards, while preventing from a significant increase in the computational complexity. 

\begin{figure}
	\centering
	\epsfig{file=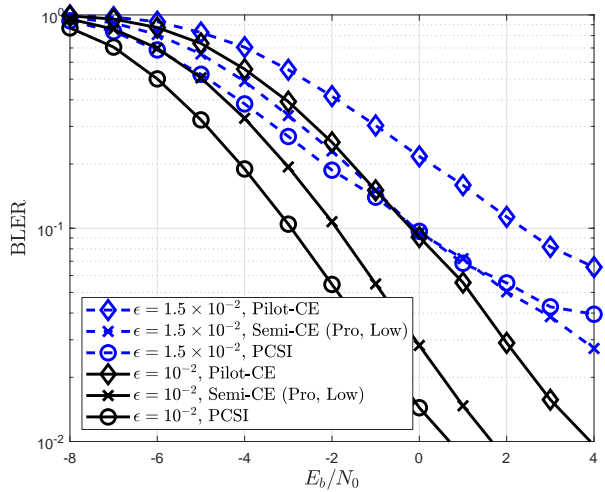, width=9cm}
	\caption{Performance comparison of various channel estimators in time-varying channels.}
	\label{FigTV}
\end{figure}

Fig. \ref{FigTV} compares the BLER of various channel estimators in time-varying channels.
To model these channels, we adopt the first-order Gaussian-Markov process in \cite{Dong2004,Kim2021} in which the channel matrix at time slot $n$ is given by 
\begin{align}\label{IV.2}
    \mathbf{H}^{(n)} &= \sqrt{1-\epsilon^{2}}\mathbf{H}^{(n-1)} + \epsilon \mathbf{E}^{(n)},
\end{align}
for $n\in \mathcal \{-{T}_{\rm{p}}{+}1,\ldots, {T}_{\rm{d}}\}$, where $\epsilon \in [0,1]$ is a temporal correlation coefficient, and $\mathbf{E}^{(n)} \in \mathbb{C}^{N_{\rm{rx}}\times N_{\rm{tx}}}$ is an i.i.d. Gaussian random matrix with zero mean and unit variance.
In this simulation, the temporal correlation coefficient is set as $\epsilon=1.5\times10^{-2}$ or $\epsilon = 10^{-2}$.
Note that {\bf PCSI} in Fig. \ref{FigTV} assumes perfect CSIR only at the {\em beginning} of data transmission (i.e., $n=1$), while it does not assume perfect channel tracking during data transmission. 
Fig. \ref{FigTV} shows that the BLER performance loss due to channel estimation error is more severe when the wireless channel varies over time because accurate channel estimation is more challenging in time-varying channels. 
In particular, when $\epsilon=1.5\times10^{-2}$, {\bf PCSI} at $n=1$ still shows severe degradation in the BLER performance if the receiver does not properly track temporal channel variations.
Unlike this, the proposed channel estimator is shown to be robust against temporal channel variations because it has a potential to track the channel variations during the first $T_{\rm u}$ time slots, by exploiting detected symbol vectors as additional pilot signals.

\section{Conclusions}
In this paper, we have proposed a semi-data-aided LMMSE channel estimator for MIMO systems.
The key idea of the proposed estimator is to selectively exploit detected symbol vectors as additional pilot signals, while optimizing this selection via RL. 
To this end, we have defined the MDP for symbol vector selection and then developed a novel RL algorithm based on the MCTS approach.  
Using simulations, we have demonstrated that the proposed channel estimator reduces the NMSE in channel estimation, while improving the BLER of the system, compared to conventional pilot-aided channel estimation.
Meanwhile, the proposed channel estimator significantly reduces communication latency for updating a channel estimate compared to conventional iterative data-aided channel estimators.
An important future research direction is to develop a semi-data-aided channel estimator for wideband systems by modifying both the MDP and the proposed RL algorithm. 
It would also be interesting to develop a semi-data-aided channel estimator for time-varying systems by properly defining the reward function of the MDP to consider the effect of temporal channel variations.

\appendix

Let $\mathbf{C}_{e}\left(\mathrm{S}_{n}\right)=\mathbb{E}\big[(\hat{\bf h}_{r}\left(\mathrm{S}_{n}\right) - \mathbf{h}_{r})(\mathbf{\hat{h}}_{r}\left(\mathrm{S}_{n}\right) - \mathbf{h}_{r})^{H}\big]$ be the error covariance matrix between ${\bf h}_r$ and $\hat{\bf h}_{r}\left(\mathrm{S}_{n}\right)$, where $\hat{\bf h}_{r}\left(\mathrm{S}_{n}\right)$ are the $r$-th row of $\hat{\bf H}\left(\mathrm{S}_{n}\right) $.
This covariance matrix does not depend on the index of a receive antenna because the channel and the noise distributions are assumed to be equal across different receive antennas.
Therefore, the MSE of the channel estimate at the state $\mathrm{S}_{n}\in\mathcal{S}_{n}$ is given by
\begin{align}\label{A.1}
\mathbb{E}\big[\|\mathbf{\hat{H}}\left(\mathrm{S}_{n}\right) - \mathbf{H} \|_{\rm F}^{2}\big]=N_{\rm rx}\text{Tr}\left[ \mathbf{C}_{e}\left(\mathrm{S}_{n}\right)  \right].
\end{align}
Utilizing this fact, the reward function in \eqref{III.B.4-2} associated with the state transition from $\mathrm{S}_{n}\in \mathcal{S}_{n}$ to $\mathrm{S}_{n+1}\in \mathcal{S}_{n+1}$ is computed as
\begin{align}\label{A.1-1}
\mathsf{R}\left(\mathrm{S}_{n},\mathrm{S}_{n+1}\right)
=\text{Tr}\left[ \mathbf{C}_{e}\left(\mathrm{S}_{n}\right) -\mathbf{C}_{e}\left(\mathrm{S}_{n+1}\right)  \right].
\end{align}
Meanwhile, when $\mathsf{\tilde{U}}\left(\mathrm{S}_{n}|a\right) \in \mathcal{S}_{n+1}$, the future reward in \eqref{III.B.5-3} can be expressed by exploiting \eqref{eq:tree_pro}, \eqref{eq:rollout_pro}, and \eqref{III.C.2-4} as
\begin{align}\label{A.2}
\mathsf{{V}}^{\star}\left(\mathsf{\tilde{U}}\left(\mathrm{S}_{n}|a\right) \right)&=\sum_{\mathbf{a}^{\rm{t}} \in \mathcal{A}^{N}}\omega_n^{\rm t}(\mathbf{a}^{\rm t})\Bigg[\sum_{m=1}^{N}\mathsf{R}\left(\mathsf{\tilde{U}}\left(\mathrm{S}_{n}|[a,\mathbf{a}_{1:m-1}^{\rm{t}}]\right),\mathsf{\tilde{U}}\left(\mathrm{S}_{n}|[a,\mathbf{a}_{1:m}^{\rm{t}}]\right)\right)\nonumber\\
& ~~~+\sum_{l=1}^{T_{\rm{u}}-n-N}
\mathsf{R}\left(\mathsf{\tilde{U}}\left(\mathrm{S}_{n}|[a,\mathbf{a}^{\rm{t}},\mathbf{a}_{1:l-1}^{\rm{r}}]\right),\mathsf{\tilde{U}}\left(\mathrm{S}_{n}|[a,\mathbf{a}^{\rm{t}},\mathbf{a}_{1:l}^{\rm{r}}]\right)\right)\Bigg] .
\end{align}
where ${\bf a}_{1:l}=[a_1,\cdots,a_l]$ is a sub-vector of ${\bf a}$ when $l\leq m$.
We assume that $\mathbf{a}_{1:0}^{\rm t}$ is the empty set with a slight abuse of notation.
By applying \eqref{A.1-1} and \eqref{A.2} into \eqref{III.B.5-2} and \eqref{III.B.5-3}, the Q-value is obtained as
\begin{align}\label{A.2-1}
\mathsf{{Q}}\left(\mathrm{S}_{n},a\right)&=\text{Tr}\left[\mathbf{C}_{e}\left(\mathrm{S}_{n}\right)-\sum_{\mathbf{a}^{\rm{t}} \in \mathcal{A}^{N}}\omega_n^{\rm t}(\mathbf{a}^{\rm t})  \mathbf{C}_{e}\left(\mathsf{\tilde{U}}\left(\mathrm{S}_{n}|[a,\mathbf{a}^{\rm{t}},\mathbf{a}^{\rm{r}}]\right)\right) \right].
\end{align}
Then the optimal policy in \eqref{III.B.5-1} is expressed as
\begin{align}
\pi^{\star}\left(\mathrm{S}_{n}\right)
&=\argmax_{a\in\{0,1\}} \mathsf{Q}\left(\mathrm{S}_{n},a\right) \nonumber\\
&=\mathbb{I}\left[ \mathsf{Q}\left(\mathrm{S}_{n},1\right) - \mathsf{Q}\left(\mathrm{S}_{n},0\right) \geq 0 \right] \nonumber \\
&=\mathbb{I}\left[ \sum_{~\mathbf{a}^{\rm{t}}\in\mathcal{A}^{N}} 
\omega_n^{\rm t}(\mathbf{a}^{\rm t})\Delta_{n}(\mathbf{a}) \geq 0~ \right], \label{A.3}
\end{align}
where 
\begin{align}
    \Delta_{n}(\mathbf{a})&=\text{Tr}\big[\mathbf{C}_{e}\big(\mathsf{\tilde{U}}\left(\mathrm{S}_{n}|[0,\mathbf{a}]\right)\big)-\mathbf{C}_{e}\big(\mathsf{\tilde{U}}\left(\mathrm{S}_{n}|[1,\mathbf{a}]\right)\big)\big], \label{A.Delta} \\ 
    \omega_n^{\rm t}(\mathbf{a}^{\rm t}) &=   \prod\limits_{l=1}^{|{\bf a}|}~ \pi^{\rm{t}}\left(\mathsf{\tilde{U}}\left(\mathrm{S}_{n}|[a,\mathbf{a}_{1:l-1}^{\rm t}]\right),a_{l}\right) \nonumber \\
    &=\prod\limits_{l=1}^{|{\bf a}|}~ (\hat{\theta}_{\hat{k}_{n+l}}[n+l])^{a_{l}}\big(1-\hat{\theta}_{\hat{k}_{n+l}}[n+l]\big)^{1-a_{l}}. \label{A.omega}
\end{align}

The remaining task is to characterize $\Delta_{n}(\mathbf{a})$ in  \eqref{A.Delta}.
From \eqref{III.B.3-2} and \eqref{III.C.2-4}, the virtual state $\mathsf{\tilde{U}}\left(\mathrm{S}_{n}|[a,\mathbf{a}]\right)\in \mathcal{S}_{n+m}$ is characterized as
\begin{align}\label{A.4-0}
	\mathsf{\tilde{U}}\left(\mathrm{S}_{n}|[a,{\bf a}]\right)
	&= (\mathbf{X}_{n+m},\mathbf{\hat{X}}_{n+m},\mathbf{a}_{n+m})\nonumber\\
	&=\begin{cases}  \left(\big[{\bf X}_{n},\mathbf{x}_{k_{n}},\tilde{\bf X}_n({\bf a})\big],
	\big[\hat{\bf X}_{n},\mathbf{\hat{x}}[n],\hat{\bf X}_n({\bf a})\big],
	[{\bf a}_n,1, {\bf a}]\right) & a=1\\
	\left(\big[{\bf X}_{n},\tilde{\bf X}_n({\bf a})\big], 
	\big[\hat{\bf X}_{n},\hat{\bf X}_n({\bf a})\big],
	[{\bf a}_n,0, {\bf a}]\right), & a=0.  \end{cases}
\end{align}
Therefore, from \eqref{II.A.2} and \eqref{A.4-0}, the distribution of
$\mathbf{\bar{y}}_{r}^{H}\big(\mathsf{\tilde{U}}\left(\mathrm{S}_{n}|[a,\mathbf{a}]\right)\big)$ is given by
\begin{align}\label{A.4}
\mathbf{\bar{y}}_{r}^{H}\big(\mathsf{\tilde{U}}\left(\mathrm{S}_{n}|[a,\mathbf{a}]\right)\big)
&\sim  \mathcal{CN}\left(\mathbf{0}_{\|\mathbf{a}_{n+m}\|_{0}} ,\mathbf{X}_{n+m}^{H} \mathbf{X}_{n+m} +\sigma^{2}\mathbf{I}_{\|\mathbf{a}_{n+m}\|_{0}}\right).
\end{align}
Using this fact, the error covariance matrix in \eqref{A.Delta} is expressed as
\begin{align}\label{A.5}
&\mathbf{C}_{e}\big(\mathsf{\tilde{U}}\left(\mathrm{S}_{n}|[a,\mathbf{a}]\right)\big) =\sigma^{2}\mathbf{Q}_{n}([a,\mathbf{a}])-\sigma^{4}\mathbf{Q}_{n}^{2}([a,\mathbf{a}])+\mathbf{Q}_{n}([a,\mathbf{a}])\mathbf{D}_{n}([a,\mathbf{a}])\mathbf{D}_{n}^{H}([a,\mathbf{a}])\mathbf{Q}_{n}([a,\mathbf{a}]),
\end{align}
where
\begin{align}
\mathbf{Q}_{n}([a,\mathbf{a}])&=\left(\mathbf{\hat{X}}_{n+m}\mathbf{\hat{X}}_{n+m}^{H}+\sigma^{2}\mathbf{I}_{N_{\rm{tx}}}\right)^{-1}\nonumber\\
&\overset{(a)}{=}\begin{cases} \big(\mathbf{\hat{X}}_{n}\mathbf{\hat{X}}_{n}^{H}+\mathbf{\hat{X}}_{n}(\mathbf{a})\mathbf{\hat{X}}_{n}^{H}(\mathbf{a})+\sigma^{2}\mathbf{I}_{N_{\rm{tx}}}\big)^{-1}, & a=0, \\ \left(\mathbf{Q}_{n}^{-1}([0,\mathbf{a}]) + \mathbf{\hat{x}}[n]\mathbf{\hat{x}}^{H}[n]\right)^{-1}, & a=1, \end{cases}\nonumber\\
\mathbf{D}_{n}([a,\mathbf{a}])&=\mathbf{\hat{X}}_{n+m}\left(\mathbf{\hat{X}}_{n+m}-\mathbf{X}_{n+m}\right)^{H} +\sigma^{2}\mathbf{I}_{N_{\rm{tx}}}\nonumber\\
&\overset{(b)}{=}\begin{cases} \mathbf{\hat{X}}_{n}\big(\mathbf{\hat{X}}_{n}-\mathbf{X}_{n}\big)^{H}+\mathbf{\hat{X}}_{n}(\mathbf{a})\big(\mathbf{\hat{X}}_{n}(\mathbf{a})-\mathbf{\tilde{X}}_{n}(\mathbf{a})\big)^{H} +\sigma^{2}\mathbf{I}_{N_{\rm{tx}}},&a=0, \\ \mathbf{D}_{n}([0,\mathbf{a}])+\mathbf{\hat{x}}[n]\left( \mathbf{\hat{x}}[n] -\mathbf{\tilde{x}}[n]\right)^{H}, &a=1, \end{cases}\nonumber
\end{align}
and both (a) and (b) come from \eqref{A.4-0}.
By using the error covariance matrix in \eqref{A.5}, $\Delta_{n}(\mathbf{a})$ in  \eqref{A.Delta} is rewritten as
\begin{align}\label{A.8}
&\Delta_{n}(\mathbf{a})
=\sigma^{2}\underbrace{\text{Tr}\left[\mathbf{Q}_{n}([0,\mathbf{a}])-\mathbf{Q}_{n}([1,\mathbf{a}])\right]}_{=A_n({\bf a})}-\sigma^{4}\underbrace{\text{Tr}\left[\mathbf{Q}_{n}^{2}([0,\mathbf{a}])-\mathbf{Q}_{n}^{2}([1,\mathbf{a}])\right]}_{=B_n({\bf a})}\nonumber\\
&+\underbrace{\text{Tr}\left[\mathbf{Q}_{n}([0,\mathbf{a}])\mathbf{D}_{n}([0,\mathbf{a}])\mathbf{D}_{n}^{H}([0,\mathbf{a}])\mathbf{Q}_{n}([0,\mathbf{a}])-\mathbf{Q}_{n}([1,\mathbf{a}])\mathbf{D}_{n}([1,\mathbf{a}])\mathbf{D}_{n}^{H}([1,\mathbf{a}])\mathbf{Q}_{n}([1,\mathbf{a}])\right]}_{=C_n({\bf a})}.
\end{align}
By the matrix inversion lemma, the matrix $\mathbf{Q}_{n}([1,\mathbf{a}])$ is rewritten as
\begin{align}\label{A.6}
\mathbf{Q}_{n}([1,\mathbf{a}])
&=\mathbf{Q}_{n}([0,\mathbf{a}]) -\frac{\mathbf{Q}_{n}([0,\mathbf{a}])\mathbf{\hat{x}}[n]\mathbf{\hat{x}}^{H}[n]\mathbf{Q}_{n}([0,\mathbf{a}])}{1+\mathbf{\hat{x}}^{H}[n]\mathbf{Q}_{n}([0,\mathbf{a}])\mathbf{\hat{x}}[n]}.
\end{align}
From \eqref{A.6}, the first term of the right-hand-side (RHS) of \eqref{A.8} is expressed as
\begin{align}\label{A.11-0}
A_n({\bf a})
&= \text{Tr}\left[ \frac{\mathbf{Q}_{n}([0,\mathbf{a}])\mathbf{\hat{x}}[n]\mathbf{\hat{x}}^{H}[n]\mathbf{Q}_{n}([0,\mathbf{a}])}{1+\mathbf{\hat{x}}^{H}[n]\mathbf{Q}_{n}([0,\mathbf{a}])\mathbf{\hat{x}}[n]} \right]
=\|\mathbf{t}_{n}(\mathbf{a})\|^{2},
\end{align}
where ${\bf t}_{n}({\bf a})=\frac{1}{\sqrt{1+\alpha_n({\bf a})}}{\bf Q}_{n}([0,{\bf a}])\mathbf{\hat{x}}[n]$ with ${\alpha}_{n}({\bf a})=\mathbf{\hat{x}}^{H}[n]{\bf Q}_{n}([0,{\bf a}])\mathbf{\hat{x}}[n]$.
The second term of the RHS of \eqref{A.8} is expressed as
\begin{align}\label{A.11}
B_n({\bf a})
&=\text{Tr}\left[\left(\mathbf{Q}_{n}([0,\mathbf{a}])-\mathbf{Q}_{n}([1,\mathbf{a}])\right)^{H}\left(\mathbf{Q}_{n}([0,\mathbf{a}])+\mathbf{Q}_{n}([1,\mathbf{a}])\right)\right]\nonumber\\
&=2\beta_{n}(\mathbf{a})\|\mathbf{t}_{n}(\mathbf{a})\|^{2}-\|\mathbf{t}_{n}(\mathbf{a})\|^{4},
\end{align}
where $\beta_{n}(\mathbf{a})=\frac{1}{\|{\bf t}_{n}({\bf a})\|^2}{\bf t}_{n}^H({\bf a}){\bf Q}_{n}([0,{\bf a}]){\bf t}_{n}({\bf a})$.
The last term of the RHS of \eqref{A.8} is computed as 
\begin{align}\label{A.9}
C_n({\bf a})
&=\text{Tr}\Big[\mathbf{Q}_{n}([0,\mathbf{a}])\mathbf{D}_{n}([0,\mathbf{a}])\mathbf{D}_{n}^{H}([0,\mathbf{a}])\mathbf{Q}_{n}([0,\mathbf{a}])\nonumber\\
&~~~~~~-\left(\mathbf{Q}_{n}([0,\mathbf{a}])-\mathbf{t}_{n}(\mathbf{a})\mathbf{t}_{n}^{H}(\mathbf{a})\right)  \big(\mathbf{D}_{n}([0,\mathbf{a}])+\mathbf{\hat{x}}[n](\mathbf{\hat{x}}[n]-\mathbf{\tilde{x}}[n])^{H}\big) \nonumber\\  
&~~~~~~\times\big(\mathbf{D}_{n}([0,\mathbf{a}])+\mathbf{\hat{x}}[n](\mathbf{\hat{x}}[n]-\mathbf{\tilde{x}}[n])^{H} \big)^{H}\left(\mathbf{Q}_{n}([0,\mathbf{a}])-\mathbf{t}_{n}(\mathbf{a})\mathbf{t}_{n}^{H}(\mathbf{a})\right)\Big]\nonumber\\
&=
\|\mathbf{t}_{n}(\mathbf{a})\|^{2}\left(\|\mathbf{v}_{n}(\mathbf{a}) \|^{2}-\|\mathbf{e}_{n}(\mathbf{a})-\mathbf{u}_{n}(\mathbf{a})+\mathbf{v}_{n}(\mathbf{a}) \|^{2}\right),
\end{align}
where ${\bf e}_n({\bf a}) = \frac{1}{\sqrt{1+\alpha_n({\bf a})}}( \mathbf{\hat{x}}[n]-\mathbf{\tilde{x}}[n])$, $
\mathbf{{v}}_{n}({\bf a})=\frac{1}{\|{\bf t}_{n}({\bf a})\|^2}{\bf D}_n^H({\bf a}) {\bf Q}_{n}({\bf a}){\bf t}_{n}({\bf a})$, and $
\mathbf{{u}}_{n}({\bf a})={\bf D}_n^H({\bf a}) {\bf t}_{n}({\bf a})$.
Applying \eqref{A.11-0}--\eqref{A.9} into \eqref{A.Delta} yields
\begin{align}\label{A.13}
\Delta_{n}(\mathbf{a})&= \|\mathbf{t}_{n}(\mathbf{a})\|^{2}\left(\sigma^{2}+\sigma^{4}\left(\|\mathbf{t}_{n}(\mathbf{a})\|^{2}- 2\beta_{n}(\mathbf{a})\right)+ \|\mathbf{v}_{n}(\mathbf{a}) \|^{2}-\|\mathbf{e}_{n}(\mathbf{a})-\mathbf{u}_{n}(\mathbf{a})+\mathbf{v}_{n}(\mathbf{a}) \|^{2}\right).
\end{align}
Finally, we obtain the result in \eqref{III.C.3-5} from \eqref{A.3} with \eqref{A.13} and \eqref{A.omega}, where $\mathbf{Q}_{n}(\mathbf{a})=\mathbf{Q}_{n}([0,\mathbf{a}])$ and $\mathbf{D}_{n}(\mathbf{a})=\mathbf{D}_{n}([0,\mathbf{a}])$.

\end{document}